\documentclass[aps]{revtex4}
\usepackage{amsmath}
\usepackage{tikz}
\usetikzlibrary{calc,decorations.pathmorphing}
\usepackage{amssymb}
\usepackage{epsfig}
\usepackage{amscd}
\usepackage{graphicx,xspace}
\usepackage{float}
\usepackage[normalem]{ulem}
\usepackage{color}
\usepackage{enumerate}

\begin{document}

\title{Conformal field theory of critical Casimir forces between
  surfaces with
  alternating boundary conditions in two dimensions}

\author{J. Dubail}
\affiliation{CNRS  \&  IJL-UMR  7198,  Universit\'e de Lorraine, F-54506 Vandoeuvre-les-Nancy,  France}
\author{R. Santachiara}
\affiliation{Laboratoire de Physique
Th\'eorique et Mod\`eles Statistiques, CNRS UMR 8626,
Universit\'e Paris-Saclay, 91405 Orsay cedex, France}
\author{T. Emig}
\affiliation{MultiScale Materials Science
for Energy and Environment, Joint MIT-CNRS Laboratory (UMI 3466),
Massachusetts Institute of Technology,  Cambridge, Massachusetts 02139, USA}
\affiliation{Laboratoire de Physique
Th\'eorique et Mod\`eles Statistiques, CNRS UMR 8626,
Universit\'e Paris-Saclay, 91405 Orsay cedex, France}

\begin{abstract}
  Systems as diverse as binary mixtures and inclusions in biological
  membranes, and many more, can be described effectively by
  interacting spins. When the critical fluctuations in these systems
  are constrained by boundary conditions, critical Casimir forces
  (CCF) emerge. Here we analyze CCF between boundaries with
  alternating boundary conditions in two dimensions, employing
  conformal field theory (CFT). After presenting the concept of
  boundary changing operators, we specifically consider two different
  boundary configurations for a strip of critical Ising spins: (I)
  alternating equi-sized domains of up and down spins on both sides of
  the strip, with a possible lateral shift, and (II) alternating
  domains of up and down spins of different size on one side and
  homogeneously fixed spins on the other side of the strip.
  Asymptotic results for the CCF at small and large distances are
  derived. We introduce a novel modified Szeg\"o formula for
  determinants of real antisymmetric block Toeplitz matrices to obtain
  the exact CCF and the corresponding scaling functions at all
  distances.  We demonstrate the existence of a surface
  Renormalization Group flow between universal force amplitudes of
  different magnitude and sign. The Casimir force can vanish at a
    stable equilibrium position that can be controlled by parameters
    of the boundary conditions. Lateral Casimir forces assume a
    universal simple cosine form at large separations.
\end{abstract}

\maketitle

\section{Introduction}

Van der Waals interactions or more generally Casimir forces are
ubiquitous in nature \cite{Parsegian:2005ly}. They originate from a
confinement or modification of fluctuations. Those can be quantum
fluctuations, as in the quantum electrodynamical (QED) Casimir
effect~\cite{Casimir:1948bh,Bordag:2009ve}, or thermal order parameter
fluctuations in the vicinity of a phase transition where correlation
lengths are large, resulting in so-called critical Casimir forces
(CCFs) between the confining boundaries
\cite{Gennes:1978qf,Krech:1994kl}.  Analogies and differences between
quantum and thermally induced forces have been reviewed in
Ref.~\cite{Gambassi:2009rw}.  The CCF is characterized by a universal
scaling function that depends on the ratio of the correlation length
and the distance between the confining elements.  This function is
determined by the universality classes of the critical medium
\cite{Diehl:1986xq}.  Its sign depends on the boundary conditions for
the order parameter at the surfaces, and hence the CCF can be
attractive or repulsive \cite{Abraham:2010,Vasilyev:2011}.
Controlling the sign of the force is important to myriad applications
in design and manipulation of micron scale devices.  Experimentally,
sign control has been achieved with judicious choice of materials in
case of QED Casimir forces~\cite{Munday:2009}, and with appropriate
boundary conditions for CCFs in binary
mixtures~\cite{Soyka:2008,Hertlein:2008,Nellen:2009}.  In QED, a
general theorem shows that there is always attraction between mirror
symmetric shapes~\cite{Kenneth:2006wn,Bachas:2007vz}.  More generally,
a theorem for Casimir forces in QED, similar to Earnshaw's theorem,
rules out the possibility of stable levitation (and consequently force
reversals) in most cases~\cite{Rahi:2010yg}. Contrary to that, the
sign of the CCF can be tailored by modifying the shape
\cite{Bimonte:2015ph} or the boundary conditions \cite{Kleban:1991ff,Kleban:1996pi} of the
confining surfaces.  For example, a classical binary mixture can be
described by an Ising model where homogeneous surfaces have a
preference for one of the two components of the mixture, corresponding
to fixed spin boundary conditions ($+$ or $-$). Depending on whether
the conditions are like ($++$ or $--$) or unlike ($+-$ or $-+$) on two
surfaces, the CCF between them is attractive or repulsive.  However,
so-called ordinary or free spin boundary conditions are difficult to
realize experimentally but can emerge due to renormalization of {\it
  inhomogeneous} conditions as we shall show below
\cite{Toldin:2013uk}.  In general, the required conditions for a sign
change remain an open problem for general shapes and boundary
conditions.

Experimentally, CCFs can be observed indirectly in wetting films of
critical fluids \cite{Nightingale:1985rm}, as has been demonstrated
close to the superfluid transition of ${}^4$He \cite{Garcia:1999} and
binary liquid mixtures \cite{Garcia:2002rc}. More recently, the CCF
between colloidal particles and a planar substrate has been measured
directly in a critical binary liquid mixture
\cite{Hertlein:2008,Soyka:2008}.  Motivated by their potential
relevance to nano-scale devices, fluctuation forces in the presence of
geometrically or chemically structured surfaces have been at the focus
recently. Chemical surface preparation allows for an adsorption
preference for the components of a binary mixture that varies along
the surface \cite{sprenger2006forces,trondle2009normal,
  trondle2010critical}. The CCF between such inhomogeneous surfaces is
determined by the effective boundary conditions at which the surfaces
``see'' each other. Due to renormalization, the effective boundary
conditions depend on the distance between the surfaces. This leads to
interesting phenomena such as cross-overs with respect to strength and
even sign of the force, a lateral force \cite{Nowakowski:2014qv} and
pattern formation among colloidal particles near non-uniform
substrates. The latter situation has been studied experimentally for
spherical colloids \cite{Soyka:2008}.  Due to the possibility that the
lipid mixtures composing biological membranes are poised at
criticality~\cite{Baumgart:2007,Veatch:2007}, it has been also
proposed that inhomogeneities on such membranes are subject to a
CCF~\cite{Machta:2012fu} which provides an example of a 2D
realisation.

Initial studies of CCFs mostly considered highly symmetric shapes and
boundary conditions since the computation of these interactions is
notoriously difficult.  Chemical surface structures (and hence
modifications of the boundary conditions) and less symmetric geometric
shapes complicate CFFs further but add also to the richness of
phenomena that can be expected. The experimentally mostly studied
critical Casimir systems belong to the Ising universality class and
hence the CCF can be extracted from numerical simulations. This has
been done for the simple film geometry with various homogeneous
boundary conditions
\cite{Vasilyev:2007sf,Hasenbusch:2010wq,Vasilyev:2013hb} and for
spherical particles \cite{Hasenbusch:2013ee}.  On the analytical side,
mostly mean field methods in combination with the Derjaguin
approximation have been used to compute CCFs for various geometries
\cite{Labbe-Laurent:2016cq}. Conformal field theory
(CFT)~\cite{belavin1984infinite,henkel2013conformal,francesco2012conformal}
opens a path to compute CCFs exactly in two dimensional systems at
criticality: Casimir forces in a strip are related to the central
charge of the CFT~\cite{Cardy:1986fu,Kleban:1991ff,Kleban:1996pi},
with appropriate modification for boundaries \cite{Cardy:1989xe}.
There are results for interactions between
circles~\cite{Machta:2012fu}, needles~\cite{Vasilyev::2012dz};
Ref.~\cite{Bimonte:2013cl} describes a general approach for any
compact shapes.  Sign changes of CCFs due to wedge like surface
structures have been reported very recently \cite{Bimonte:2015ph}.
These techniques have been applied to the Casimir interaction between
two needles immersed in a two-dimensional critical fluid of Ising
symmetry \cite{Eisenriegler:2016wq}. A general classification of the
sign of the critical Casimir force within 2D CFT's has been obtained
within a similar approach \cite{Rajabpour:rp}.

\begin{figure}[h]
	\begin{tikzpicture}
		\begin{scope}[scale=0.4]
			\draw[<->] (2.5,6.5) -- (2.5,7.5);
			\draw (1.5,7) node{$a_0$};
			\filldraw (9,7) circle (2mm);
			\filldraw (5,2) circle (2mm);
			\filldraw (3,5) circle (2mm);
			\draw (5,8.6) node{${\rm BC}_1$};
			\draw (9.3,4.5) node{${\rm BC}_2$};
			\draw (1.2,3) node{${\rm BC}_3$};
			\draw[<->] (2,1) -- (9,1);
			\draw (5.5,0.5) node{$L$};
			\draw[clip] plot[smooth cycle] coordinates{(3,5) (4,8) (9,7) (8,5) (8,4) (5,2) (2,2)};
			\draw[ultra thick] plot[smooth cycle] coordinates{(3,5) (4,8) (9,7) (8,5) (8,4) (5,2) (2,2)};			
			\foreach \x in {0,0.8,...,10} {
				\draw (0.2+\x,0) -- (0.2+\x,10);
				\draw (0,\x+0.2) -- (10,\x+0.2);
			}
		\end{scope}
		\draw (6.25,2.5) node{$L / a_0 \gg 1$};
		\draw[->] (5.5,2) -- (7,2);
		\begin{scope}[xshift=8cm,scale=0.4]
			\filldraw (9,7) circle (2mm);
			\filldraw (5,2) circle (2mm);
			\filldraw (3,5) circle (2mm);
			\draw (5,8.6) node{${\rm BC}_1$};
			\draw (9.3,4.5) node{${\rm BC}_2$};
			\draw (1.2,3) node{${\rm BC}_3$};
			\draw[clip] plot[smooth cycle] coordinates{(3,5) (4,8) (9,7) (8,5) (8,4) (5,2) (2,2)};
			\draw[ultra thick] plot[smooth cycle] coordinates{(3,5) (4,8) (9,7) (8,5) (8,4) (5,2) (2,2)};			
			\foreach \x in {0,0.2,...,10} {
				\draw (0.2+\x,0) -- (0.2+\x,10);
				\draw (0,\x+0.2) -- (10,\x+0.2);
			}
		\end{scope}
	\end{tikzpicture}
	\caption{A statistical lattice model in a domain with a given
          shape, and a given set of boundary conditions
          (${\rm BC}_1$, ${\rm BC}_2$, ${\rm BC}_3$). The dots
          indicate positions where the boundary condition changes. The
          system has a linear size $L$, and a UV cutoff $a_0$ (here
          represented as lattice spacing). Sending the UV cutoff to
          zero, while keeping the global shape fixed, one obtains a
          system that is well described by a conformal field theory
          (CFT) in the bulk, with boundary condition changing
          operators (BCC operators) located at the dots on the
          boundary. Independent of microscopic details, the
          predictions of CFT are applicable to all models that fall
          into the same universality class for $L \gg a_0$.}
	\label{fig:lattice}
\end{figure}
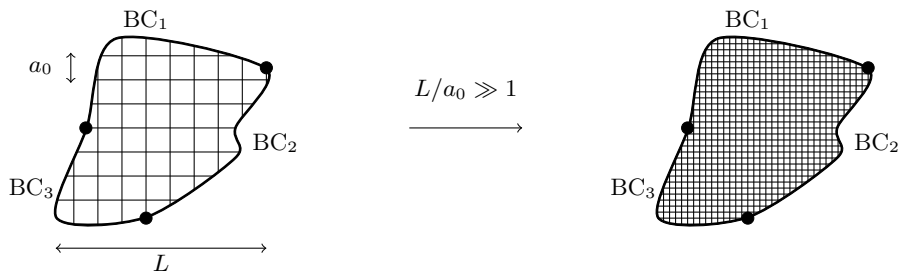

How can one describe a system with changing boundary conditions in
CFT?  Consider a bounded 2d lattice model at equilibrium. For example,
one may think of an Ising or Potts model. Imagine that the boundary
shape of the system is fixed, as well as its boundary conditions. Let
$L$ be the linear size of the system, and $a_0$ a UV cutoff, for
example the lattice spacing. Imagine that we were able to tune the UV
cutoff without affecting the global shape of the system
(Fig. \ref{fig:lattice}). Then one may consider $L/a_0$ as a large
parameter, and focus on the corresponding expansion of the free
energy, which is usually of the form
$\mathcal{F} = \alpha_{{\rm bulk}} \frac{L^2}{a_0^2} + \alpha_{{\rm
    surf.}} \frac{L}{a_0} + \dots$
The first two terms reflect the fact that the free energy is extensive
(at least for systems with sufficiently short-ranged interactions):
they scale with the bulk area and with the length of the boundary. At
criticality, it is well-known that the next term in this expansion is
universal and may be computed using the powerful machinery of
CFT,
\begin{equation}
	\mathcal{F}  =   \alpha_{\rm bulk} \frac{L^2}{a_0^2}  +   \alpha_{{\rm surf.}} \frac{L}{a_0}   +  \mathcal{F}_{{\rm CFT}} + \ldots  \, .
\end{equation}
In general, the last term can be decomposed as
\begin{equation}
	\label{eq:F_CFT}
	\mathcal{F}_{{\rm CFT}} \, = \, \zeta  \, \log \frac{L}{a_0} + \theta[{\rm shape}] \, ,
\end{equation}
with a logarithmic contribution with coefficient $\zeta$, and some
function $\theta[{\rm shape}]$ that does not depend on $L/a_0$,
but instead is a function of the shape of the system. For
example, when the system is a rectangle with sides of length $L$ and
$\ell$, then it is a function of the aspect ratio,
$\theta = \theta(\ell/L)$. For more complicated shapes, described by
more length scales $\ell_1, \ell_2, \dots$, it is a function
$\theta(\ell_1/L, \ell_2/L , \dots)$.  Importantly, both the
coefficient $\zeta$ and the function $\theta$ are
universal -- meaning that they are independent of the microscopic
details of the system (like the lattice constant or UV cutoff
$a_0$) -- and can be computed in CFT.

In the following, we focus on simply connected domains with one
boundary. In the simplest case when the boundary condition is the same
everywhere along the boundary, it is known that the CFT part
$\mathcal{F}_{{\rm CFT}} $ of the free energy is directly proportional
to the central charge $c$ of the CFT model, and is insensitive to any
other details of the theory. More precisely, $\zeta$ and
$\theta[{\rm shape}]$ are both given by $c$ times their value in the
free scalar field theory with Dirichlet boundary conditions (the normalization
convention for the central charge is such that the free scalar field
has $c=1$). The variation with the shape of the domain is
given in full generality by the Polyakov-Alvarez formula \cite{polyakov1981quantum,
alvarez1983theory}, but we will not need it in what follows. In fact, in
this paper we consider only very simple shapes, for which the CFT
part of the free energy is well-known. Instead, our main interest is
in the contributions to the free energy $\mathcal{F}_{\rm CFT}$ that
are induced by {\it changes} of the boundary conditions along the
boundary. When the boundary condition (BC) changes from ${\rm BC}_1$ to ${\rm BC}_2$, from
${\rm BC}_2$ to ${\rm BC}_3$, and so on (Fig. \ref{fig:lattice}), at points
$z_1, \ldots, z_m$ along the boundary, the partition
function, $Z = (L/a_0)^{-\zeta} e^{- \theta[{\rm shape}]}$, can be expressed as
\begin{equation}
	\label{eq:bcc_correl}
	\frac{Z}{Z_0} =  \left< \phi_{{\rm BC}_1 | {\rm BC}_2} (z_1) \phi_{{\rm BC}_2 | {\rm BC}_3} (z_2) \dots \phi_{{\rm BC}_n | {\rm BC}_1} (z_n) \right>  .
\end{equation}
Here $Z_0$ is the partition function when the BC is {\it constant} along
the boundary, and $Z$ is the partition function of the system {\it with} the BC
changes. The remarkable thing is that, since the change of BC is a local
effect, in the scaling limit it can be encoded as a local operator \cite{Cardy:1989xe},
dubbed {\it boundary condition changing} (BCC) operator. These operators are
{\it primary} operators, with a scaling dimension $h_p$ that depends on the two boundary
conditions ${\rm BC}_p$ and ${\rm BC}_{p+1}$. 
As a consequence, the joint effect of several changes of boundary
conditions must take the form of a correlation function of local
operators along the boundary. How do such changes of boundary
conditions affect the free energy of Eq.~(\ref{eq:F_CFT})?  The
insertion of BCC operators is reflected in the free energy by
\begin{eqnarray}
\label{eq:bcc_correl_2}
	\mathcal{F}_{{\rm CFT}}  \; \rightarrow \;    \mathcal{F}_{{\rm CFT}} \, - \, \log \left< \phi_{{\rm BCC},1} (z_1) \dots \phi_{{\rm BCC},n} (z_n) \right> \, ,
\end{eqnarray}
and, since correlation functions must have the scaling form
$\left< \phi_{{\rm BCC},1} (z_1) \dots \phi_{{\rm BCC},n} (z_n) \right> \,
= \, (L/a_0)^{-\sum_j h_j} g(z_1/L,\dots, z_n /L)$,
where $h_j$ is the scaling dimension of the BCC operator $ \phi_{{\rm BCC},j}$,
we see that
\begin{equation}
	\zeta \rightarrow \zeta + \sum h_j \qquad {\rm and}
        \qquad 
\theta \rightarrow \theta - \log g(z_1/L,\dots, z_n /L) \, .
\end{equation}
The calculation of arbitrary correlation functions, and thus of the
function $g$, is usually difficult. The purpose of this paper is to
exploit a few cases where this can be done exactly. A short account of
our results has been published recently \cite{Dubail:2016uo}. 

The rest of the paper is organized as follows. In the next section we
consider a general CFT on a strip with a small number of changes in
the boundary conditions, to familiarize the reader with the use of BCC operators. In
Sec.~III we introduce the model of interest: two boundary configurations of the
critical Ising model on a strip with periodically alternating boundary
conditions. In that section we derive limiting cases of the CCF from simple
arguments. Some facts on Toepliz determinants and a modified Szeg\"o
formula for the determinant of real antisymmetric block Toeplitz matrices are presented in
Sec.~IV. This modified formula allows us the compute the exact CCF for the
two boundary configurations of the Ising strip in Sec.~V. We conclude in Sec.~VI.

\section{CFT for Casimir forces across a strip with changing boundary conditions}

\begin{figure}[h]
\begin{center}
\begin{tikzpicture}[scale=0.8]
	\filldraw[gray] (-8,0) rectangle (8,2);
	\draw[thick] (-8,0) -- (8,0);
	\draw[thick] (-8,2) -- (8,2);
	\draw[<->] (5,1.8) -- (5,0.2);
	\draw (4.7,1) node{$L$};
	\draw[<->] (-7.5,-0.4) -- (7.5,-0.4);
	\draw (0,-0.7) node{$W$};
	\draw[thick, dashed] (7.5,0) -- (7.5,2);
	\draw[thick, dashed] (-7.5,0) -- (-7.5,2);
\end{tikzpicture}
\end{center}
\caption{Strip of width $L$ and length $W \gg L$. Here the boundary condition is assumed to be
 identical on both sides.}
	\label{fig:strip1}
\end{figure}
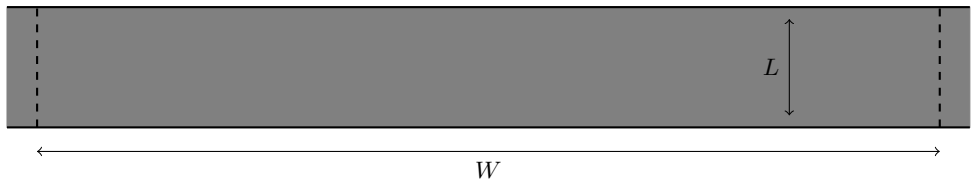

In this section, we consider a general CFT in the strip geometry, with
boundary condition changes along the surface.  Making use of the fact
that, in any CFT, two-point functions and three-point functions are
entirely fixed by conformal invariance, the CCF can be computed
explicitly, because it follows from the correlation function of BCC
operators that enters the free energy, see
Eq.(\ref{eq:bcc_correl_2}). We assume that the strip has width $L$ and
length $W$ where the latter serves as an IR cutoff that can be
considered infinity when computing the free energy density
${\cal F_{\rm CFT}}/W$ (see Fig.~{\ref{fig:strip1}}).

\subsection{Identical BCs on both sides}

It is well-know that the CFT part of the free energy of a rectangle of
size $L \times W$ for $W/L\gg 1$ is to leading order 
\begin{equation}
\label{eq: torusZ}
\mathcal{F}_{\rm CFT} =  - \frac{\pi c}{24} \frac{W}{L} + \dots
\end{equation}
where $c$ in the central charge of the CFT, and the dots stand for
corrections that vanish when $W/L \rightarrow \infty$. The critical
Casimir force per unit length between the two boundaries is then
\begin{equation}
	\label{eq: defcasimr}
	\frac{F}{W} = -\frac{d}{d L} \frac{ \mathcal{F}_{\rm CFT}  }{W} =  - \frac{\pi c}{24 L^2} \, .
\end{equation}

\subsection{Insertion of two BCC operators}

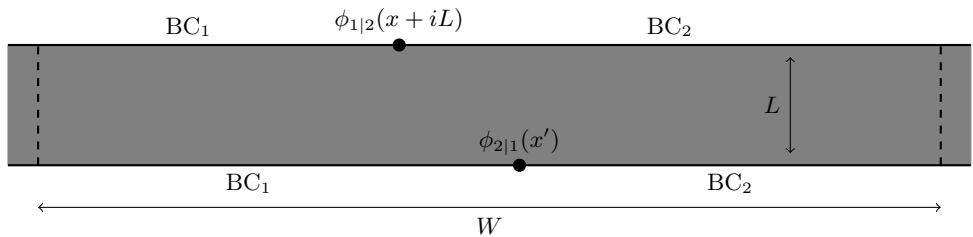
\begin{figure}[h]
\begin{center}
\begin{tikzpicture}[scale=0.8]
	\filldraw[gray] (-8,0) rectangle (8,2);
	\draw[thick] (-8,0) -- (8,0);
	\draw[thick] (-8,2) -- (8,2);
	\draw[<->] (5,1.8) -- (5,0.2);
	\draw (4.7,1) node{$L$};
	\draw[<->] (-7.5,-0.7) -- (7.5,-0.7);
	\draw (0,-1) node{$W$};
	\draw[thick, dashed] (7.5,0) -- (7.5,2);
	\draw[thick, dashed] (-7.5,0) -- (-7.5,2);
	\filldraw (-1.5,2) circle (1mm);
	\filldraw (0.5,0) circle (1mm);
	\draw (-1.5,2.4) node{$\phi_{1| 2}(x+iL)$};
	\draw (0.5,0.4) node{$\phi_{2| 1}(x')$};
	\draw (-5,2.3) node{${\rm BC}_1$};
	\draw (-4,-0.3) node{${\rm BC}_1$};
	\draw (3,2.3) node{${\rm BC}_2$};
	\draw (4,-0.3) node{${\rm BC}_2$};
\end{tikzpicture}
\end{center}
	\caption{\label{fig:strip+2bcc} A strip with two changes of boundary conditions. For better readability we write $\phi_{1| 2}$ instead of $\phi_{{\rm BC}_1 | {\rm BC}_2}$.}
\end{figure}

Now consider a strip where the boundary conditions change from
${\rm BC}_1$ to ${\rm BC}_2$ at the positions $z= x+i L$ (upper
boundary) and $z= x'$ (lower boundary), see Fig.~\ref{fig:strip+2bcc}
The two-point correlation function of the BCC operators with scaling
dimension $h$ is fixed by conformal invariance,
\begin{equation}
\label{eq: 2point}
	\left< \phi_{{\rm BC}_1| {\rm BC}_2} (x+ i L) \, \phi_{{\rm BC}_2|{\rm BC}_1} (x') \right>  =
        \frac{1}{\left| \frac{2L}{\pi}  \sinh \frac{\pi (x-x' + i
              L)}{2L} \right|^{2h}} \, .
\end{equation}
Using Eq.~\eqref{eq:bcc_correl_2}, the free energy becomes
\begin{eqnarray}
\label{eq:2_bcc_energy}
	\mathcal{F}_{\rm CFT} &=&  -\frac{\pi c}{24} \frac{W}{L}
                                  + 2 h  \, \log \left| \frac{2L}{\pi}
\cosh \frac{\pi (x-x')}{2L} \right| \, .
\end{eqnarray}
There are some limiting cases that are of primary relevance to the rest
of this paper. First, take $x \rightarrow -W/2$ and
$x' \rightarrow W/2$. This corresponds to boundary condition
${\rm BC}_2$ on the upper surface, and ${\rm BC}_1$ on the lower
surface. In that limit the free energy per unit length is to leading order
\begin{eqnarray}
\label{eq:2_bcc_energy_2}
	\frac{\mathcal{F}_{\rm CFT}}{W} &=&  - \left(\frac{c}{24} - h\right) \frac{\pi}{L} \, ,
\end{eqnarray}
where we have ignored logarithmic corrections. From this result one
observes that if the boundary condition is different on both sides,
then the critical Casimir force becomes repulsive if $h>c/24$.
Interestingly, a strip with one change of boundary conditions on the
upper boundary and constant boundary conditions on the lower boundary
has the free energy of Eq.~\eqref{eq:2_bcc_energy_2} with $h$ replaced
by $h/2$.  This can be easily seen by keeping $x$ finite and taking
$x'\to W/2 \gg x$.  Hence, like and unlike boundary conditions
contribute with equal weight to the total interaction energy between
the surfaces (addivity, see also below).

For distances $L$ large compared to $|x-x'|$ (but still small compared
to the length $W$), the second term in Eq.~\eqref{eq:2_bcc_energy} has
a leading logarithmic dependence on $L$ but is independent of $W$ so
that it does not contribute to $\mathcal{F}_{\rm CFT}/W$ in the thermodynamic limit
$W\to\infty$. Hence, in that limit the strip appears like a strip with
equal boundary conditions on both sides, and the free energy per unit
length is given by Eq.~\eqref{eq:2_bcc_energy_2} with $h=0$. In
general, a {\it finite} number of boundary changes with a {\it finite}
distance between them does not change the free energy per unit length
for asymptotically large $L$ in the thermodynamic limit.

Another limiting case of interest corresponds to short distance,
$L \ll |x-x'|$. In that limit one easily sees that the free energy per
unit length becomes to leading order
\begin{eqnarray}
	\frac{\mathcal{F}_{\rm CFT}}{W} &=& - \frac{\pi c}{24 L} + \frac{\pi h}{L} \frac{|x'-x|}{W }\, .
\end{eqnarray}
This case illustrates the {\it additivity} in short distance limit: The second term
contributes to the Casimir potential $\pi h/L$ only over the fraction $|x'-x|/W$
of the boundaries where they have different boundary conditions. 
This dependence of the free energy on the position where the BCs
change gives rise to a lateral CCF per unit length, determined by
\begin{equation}
  \label{eq:two_bcc_lateral}
  \frac{F_{\rm lat}}{W} = -\frac{d ({\cal F}_{\rm CFT}/W)}{d x'} = -\, {\rm sgn}\,(x'-x)
  \frac{\pi h}{WL} \, . 
\end{equation}
This lateral force is constant along the boundaries (up to a sign) and
tends to align the two boundaries at $x=x'$ so that
there are no different boundary conditions facing each other. Of
course, for $W/L\to\infty$ this force vanishes since the free energy
is extensive in $W$. In that limit, a finite lateral force per unit
length could be achieved by a finite {\it density} of positions where
the boundary conditions change. This situation shall be considered in
Sec.~III.

\subsection{Insertion of three BCC operators}

\begin{figure}
\begin{center}
\begin{tikzpicture}[scale=0.8]
	\filldraw[gray] (-8,0) rectangle (8,2);
	\draw[thick] (-8,0) -- (8,0);
	\draw[thick] (-8,2) -- (8,2);
	\draw[<->] (5,1.8) -- (5,0.2);
	\draw (4.7,1) node{$L$};
	\draw[<->] (-7.5,-0.7) -- (7.5,-0.7);
	\draw (0,-1.) node{$W$};
	\draw[thick, dashed] (7.5,0) -- (7.5,2);
	\draw[thick, dashed] (-7.5,0) -- (-7.5,2);
	\filldraw (-1.5,2) circle (1mm);
	\filldraw (0.5,0) circle (1mm);
	\filldraw (2.5,2) circle (1mm);
	\draw (-1.5,2.4) node{$\phi_{1|2}(w_1)$};
	\draw (2.5,2.4) node{$\phi_{2|3}(w_2)$};
	\draw (0.5,0.4) node{$\phi_{3|1}(w_3)$};
	\draw (-7,2.3) node{${\rm BC}_1$};
	\draw (0,2.3) node{${\rm BC}_2$};
	\draw (6,-0.3) node{${\rm BC}_3$};
	\draw (-6,-0.3) node{${\rm BC}_1$};
	\draw (7,2.3) node{${\rm BC}_3$};
\end{tikzpicture}
\end{center}
\caption{\label{fig:strip+3bcc} A strip with three changes of boundary
  conditions.}
\end{figure}
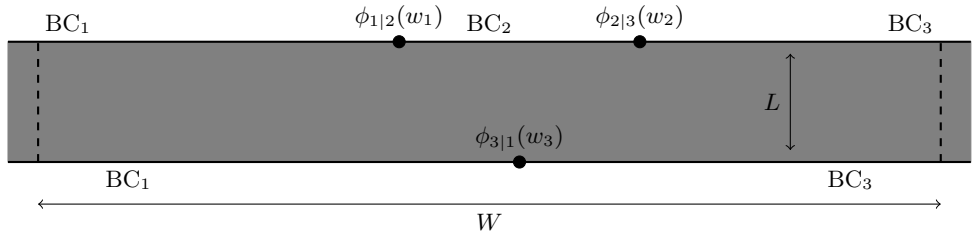
Next, we consider the situation depicted in Fig.~\ref{fig:strip+3bcc},
which is a strip with three BCC operators on the boundaries,
separating three different boundary conditions. It is well-known that the three-point correlation
function is entirely fixed by
conformal invariance \cite{henkel2013conformal,francesco2012conformal}. For three BCC operators on the boundary of the
strip, it is given by 
\begin{eqnarray}
\label{eq:3point}
	 \left< \phi_{1|2}(w_1) \phi_{2|3}(w_2) \phi_{3|1}(w_3)
  \right>  & = &  
\left(\frac{2L}{\pi}\right)^{-h_{1|2}-h_{2|3}-h_{3|1}}  
\left| \sinh \frac{\pi(w_1-w_2)}{2L} \right|^{-h_{1|2}-h_{2|3}+h_{3|1}} \\
\nonumber&&	    \quad \times \;
	 \left| \sinh \frac{\pi(w_2-w_3)}{2L} \right|^{+h_{1|2}-h_{2|3}-h_{3|1}}
  \left| \sinh \frac{\pi(w_3-w_1)}{2L} \right|^{-h_{1|2}+h_{2|3}-h_{3|1}} \, .
\end{eqnarray}
For better readability, we have replaced $\phi_{{\rm BC}_1|{\rm BC}_2}$ by $\phi_{1| 2}$; the scaling dimension of the latter operator is $h_{1|2}$. Formula (\ref{eq:3point}) easily follows from combining the conformal map $w \mapsto e^{\frac{\pi}{L} w}$, that sends the strip to the upper half-plane, with the well-known formula for the three-point function in the latter domain \cite{henkel2013conformal,francesco2012conformal}. 

Again, it is interesting to consider the limit of short distances $L$
(compared to the distances between changes in the boundary
conditions). To study this limit we set $w_1=x_1+iL$, $w_2=x_2+iL$,
$w_3=x_3$ and assume that $x_1<x_2$. The free energy can be obtained
again from Eq.~\eqref{eq:bcc_correl_2}. The result depends on the
position of $x_3$ on the lower boundary relative to $x_1$ and $x_2$ on
the upper boundary. We find for the free energy per unit length
\begin{equation}
  \label{eq:free_energy_3bcc}
  \frac{{\cal F}_{\rm CFT}}{W} = -\frac{\pi c}{24 L} + \left\{
\begin{array}{ll}
\frac{\pi}{W L} \left[ h_{2|3} (x_2 - x_1) + h_{3|1} (x_1 - x_3) \right] & {\rm for }\,\,
                                                       x_3 < x_1 <x_2 \\[1em]
\frac{\pi}{W L} \left[ h_{1|2} (x_3 - x_1) + h_{2|3} (x_2 - x_3) \right] & {\rm for }\,\,
                                                       x_1 < x_3 <x_2 \\[1em]
\frac{\pi}{W L} \left[ h_{1|2} (x_2 - x_1) + h_{3|1} (x_3 - x_2) \right] & {\rm for }\,\,
                                                       x_1 < x_2 <x_3 
\end{array}
\right. \, .
\end{equation}
Since the scaling dimensions $h_{j|k}$ are all non negative, the part of
the free energy that depends on the positions $x_j$ where the boundary
conditions change is always positive and hence all boundary changes
cost energy. To see which configuration minimizes the free energy when
we allow the lower boundary and hence $x_3$ to move laterally, we
consider the lateral force
\begin{equation}
   \label{eq:three_bcc_lateral}
  \frac{F_{\rm lat}}{W} = -\frac{d ({\cal F}_{\rm CFT}/W)}{d x_3} =
- \frac{\pi}{WL} \times \left\{
\begin{array}{ll}
-h_{3|1} & {\rm for }\,\,   x_3 < x_1 <x_2 \\[1em]
h_{1|2} - h_{2|3} & {\rm for }\,\, x_1 < x_3 <x_2 \\[1em]
h_{3|1} & {\rm for }\,\,  x_1 < x_2 <x_3 
\end{array}
\right. \, .
\end{equation}
In general, the lateral force is discontinuous at the points where the
boundary conditions change. It points to the center region between
$x_1$ and $x_2$ when $x_3$ is located outside that region so that the
lower boundary tends to move $x_3$ towards $x_1$ or $x_2$. This is
expected since otherwise there could be an arbitrarily large region
where the boundary conditions BC$_1$ and BC$_3$ face each other, which
would  cost an extensive energy. When $x_3$ is located in the
central region between $x_1$ and $x_2$, the lateral force tends to align $x_3$
with $x_1$ when $h_{1|2} > h_{2|3}$ and with $x_2$ when  $h_{1|2} <
h_{2|3}$ (If $h_{1|2} = h_{2|3}$ then there is no lateral force in
that center region).  So the stable point of minimal energy is either
$x_3=x_1$ or $x_3=x_2$, depending on the relative magnitude of 
$h_{1|2}$ and $h_{2|3}$. Again, this is expected physically as the
scaling dimensions $h_{j|k}$ measure the energy cost associated with
two different boundary conditions facing each other. So if  $h_{1|2} > h_{2|3}$
it is more expensive to have BC$_1$ and BC$_2$ opposite each other
than  BC$_2$ and BC$_3$ and hence $x_3=x_1$ minimizes the energy.

\section{Ising universality class with periodically alternating
  boundary conditions: general arguments and asymptotic limits}
\label{Ising_uc}

Any critical system in the 2d Ising universality class is described by
a CFT with central charge $c=1/2$. It admits three type of conformally
invariant boundary conditions, corresponding to fixed ($+$ or $-$)
spins and free (f) spins along the boundary. The BCC operators
implementing the change $(+,-)$ and $(+,f)$ or $(-,f)$ have dimension
$h_{(+,-)}=1/2$ and $h_{(\pm,f)}=1/16$ respectively
\cite{Cardy:1989xe}. We consider again the strip geometry. Of course, the
formulas of the previous section apply, but we would like to
generalize them to an {\it extensive} number of BCC operators. We are
interested in computing the critical Casimir forces between surfaces
with periodic chemical patterns. Since these chemical structures
realize different boundary conditions for the Ising spins when the
model is applied to binary mixtures, here we 
focus on boundary conditions which alternate
periodically. In contrast with usual boundary CFT approaches, this
problem requires to consider a {\it finite density} of BCC operators.
We consider the two following configurations:
\begin{itemize}
\item Configuration I: A strip of width $L$ and length $W\gg L$ with
  periodically alternating fixed spin boundary conditions on both
  edges with periodicity $a$ and lateral shift $\delta$, as
  illustrated in Fig.~\ref{fig:conf_I}. The positions of the $2N$ BCC
  operators are ($w= x+i y$):
\begin{equation}
	w_{2j} = j a + iL , \qquad \qquad w_{2j+1} = j a+\delta \,
        ,\qquad j=1,\ldots, N =W/a .
\label{eq:confI}
\end{equation}
For large $W$ or $N$, we are interested in the CFT part of the free
energy per unit length, $\mathcal{F}_{\rm CFT}/W$, as a function
of $L$ and of the two dimensionless parameters $a/L$ and $\delta/L$.

\begin{figure}
\begin{center}
\includegraphics[scale=0.8]{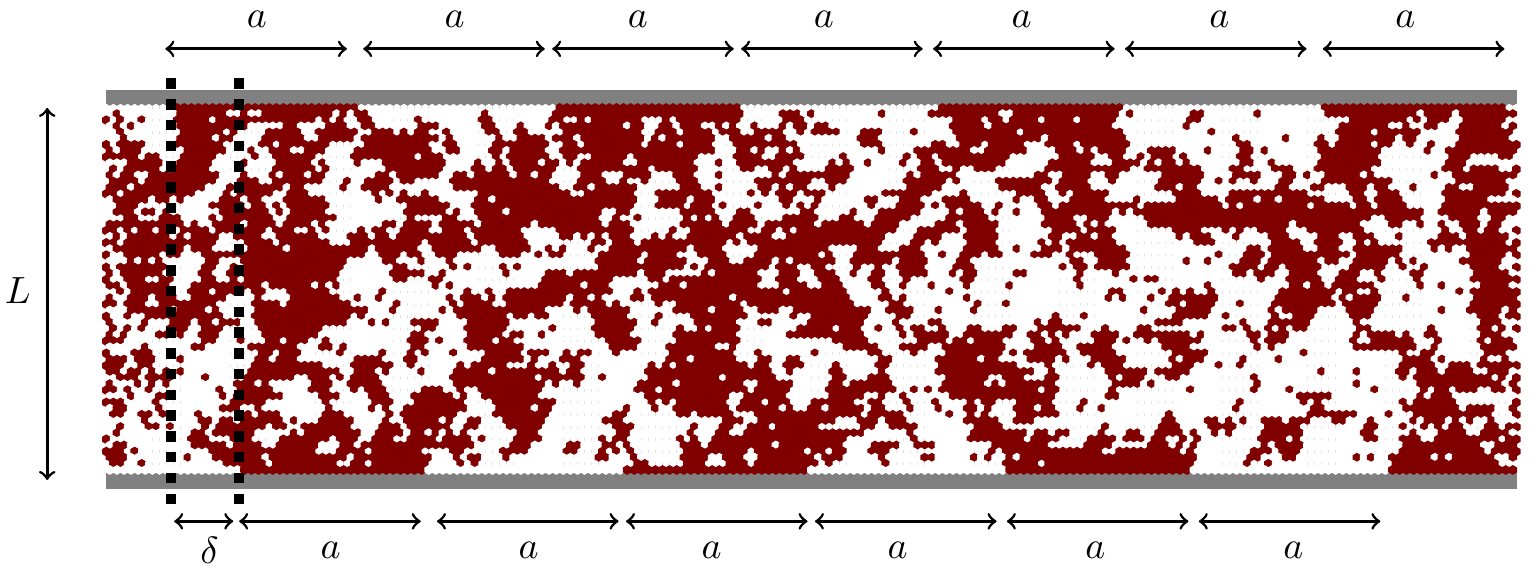}
\end{center}
\caption{\label{fig:conf_I} Ising strip with boundary configuration I.}
\end{figure}
\begin{figure}
\begin{center}
\includegraphics[scale=0.8]{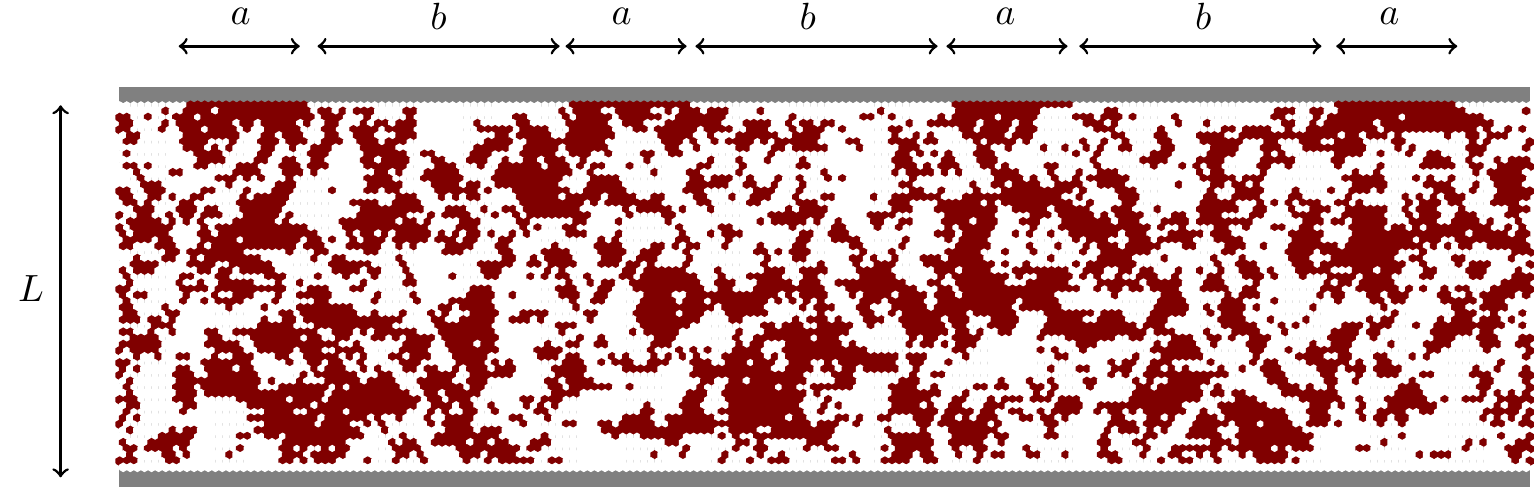}
\end{center}
\caption{\label{fig:conf_II} Ising strip with boundary configuration II.}
\end{figure}

\item Configuration II: A strip of width $L$ and length $W\gg L$ with
  homogeneous $+$ spins on the lower boundary and alternating domains
  of $−$ and $+$ spins of different lengths $a$ and $b$, respectively,
  on the upper boundary, see Fig.~\ref{fig:conf_II}.  The positions of
  the $2N$ BCC operators are
\begin{equation}
	w_{2j} = j\; (a+b), \qquad \qquad w_{2j+1} = j (a+ b)+a\, ,
        \qquad j=1, \ldots, N = W/(a+b) \, .
	\label{eq:confII}
\end{equation}
Now the CFT part of the free energy per unit length,
$\mathcal{F}_{\rm CFT}/W$, is a function of $L$, and of the
dimensionless ratios $a/L$ and $b/L$.
\end{itemize}

\subsection{Casimir interaction for small $L \ll a,b,\delta$}

When $L$ is much smaller than the separations between boundary
changes, we can use the additivity property demonstrated in the
previous section. 

\subsubsection{Configuration I}
\label{sec:smallLI}

In the short distance limit one has a very narrow strip, with segments
that have ($+$,$-$) boundary conditions on either sides over a fraction $\delta/a$ of the system, and ($+$,$+$)/($-$,$-$) boundary conditions
over a fraction $1-\delta/a$. Hence, additivity yields 
\begin{eqnarray}
\label{eq:smallLI}
\nonumber \frac{{\cal F}_{\rm CFT}}{W} &= &  
 - \frac{\pi c}{24\, L} \left(1- \frac{\delta}{a} \right)  -
                                        \frac{\pi }{L}
                                        \left(\frac{c}{24}  -
                                        h_{(+,-)} \right)
                                        \frac{\delta}{a}  
 + \ldots \\
	& = &  - \frac{\pi}{48 \,L} + \frac{\pi}{2 \, L} \frac{\delta}{a} + \ldots \, ,
\end{eqnarray}
where we used $c=1/2$ and $h_{(+,-)} =1/2$. There is also a
logarithmic contribution to $\mathcal{F}_{\rm CFT}$, which reads
$ \frac{2 W}{a} h_{(+,-)} \log (L/a_0)$ with $a_0$ a UV cutoff that
does not appear in the force after taking the derivative with respect
to $L$. However the latter is subleading in the regime $L \ll a,b$, so
we dropped it.

\subsubsection{Configuration II}

Now the narrow strip has segments 
($+$,$-$) boundary conditions over a fraction  $a/(a+b)$ of the system, and ($+$,$+$)
boundary conditions over a fraction $b/(a+b)$, yielding 
\begin{eqnarray}
\label{eq:II small L}
\nonumber \frac{{\cal F}_{CFT}}{W} &= &   -\frac{\pi c}{24\, L} \frac{b}{a+b}
 -  \frac{\pi }{L} \left(\frac{c}{24}  - h_{(+,-)} \right)
                                              \frac{a}{a+b}   + \ldots
  \\
& = & - \frac{\pi}{48 \,L} \frac{b-23 a}{a+b} + \ldots,
\end{eqnarray}
where we again used additivity. Interestingly, the force changes sign
depending on the relative amount of $+$ and $-$ spins on the
boundaries, and it vanishes to leading order for $a=b/23$. As in the
case of Configuration I, we dropped the logarithmic term, which is
subleading.

\subsection{Asymptotic Casimir interaction for large $L\gg a,b,\delta$
  from renormalization group arguments}

\subsubsection{Configuration I}
\label{sec:bigLI}

When $L \gg a$, it is useful to look at each of the two boundaries of
the strip from a coarse-grained perspective. Contrary to homogeneous
boundary conditions, in this configuration the BCC operators break the
scale invariance at the boundaries, because they are associated with
the length scale $a$. However, at length scales much larger than $a$,
one should be able to regard the coarse-grained boundary as
homogeneous, with some new effective boundary condition.
The effective boundary condition must be a Renormalization Group (RG)
fixed point \cite{diehl1997theory}.
To calculate
the free energy for Configuration I, it is not necessary to know what RG
fixed point this is exactly, because it clearly has to be the same boundary condition
on both sides,
so it is given by Eq. (\ref{eq: torusZ}), with $c=1/2$,
\begin{equation}
\frac{\mathcal{F}_{CFT}}{W} =- \frac{\pi }{48} \frac{1}{L} \, .
\end{equation}
We shall see in Sec.~V that this result is indeed recovered from
the exact solution of the model.

\subsubsection{Configuration II}

Repeating the same RG argument as for configuration I, we
now need to have a finer understanding of the boundary conditions
that are RG fixed points. For the Ising universality class,
they are expected to be one of the three known conformally invariant
boundary conditions \cite{Cardy:1989xe}: free (f), or fixed ($+$), ($-$)
boundary conditions. We thus need to distinguish the following
cases. When $a=b$, the effective boundary
condition at large scale should be the free (f)
boundary condition. Indeed, since in that case the proportion of $+$ and $-$ spins along
the boundary is the same, the effective boundary condition must enjoy
$\mathbb{Z}_2$-invariance, which rules out fixed boundary
conditions. However, when $b>a$, there are more $+$ spins
than $-$ spins on the upper boundary, so the $\mathbb{Z}_2$-symmetry
is broken, and one expects the system to renormalize towards fixed
($+$) boundary conditions. Similarly, for $a>b$, one should obtain
fixed ($-$) boundary conditions.  Since the lower boundary has
homogeneous ($+$) boundary conditions, the CFT free energy must
behave asymptotically for $W \gg L\gg a,b$ as
\begin{equation}
\label{eq: II big L}
\frac{{\cal F}_{CFT}}{W} =\left\{ \begin{array}{lll}  
- \frac{\pi}{L} \frac{c}{24} = 
- \frac{\pi }{L} \frac{1}{48}  & \text{if} \quad b>a \quad &  (+,+) \;
                                                             {\rm BC}
                                                             \, ,\\[1em]
- \frac{\pi }{L} \left(\frac{c}{24} - h_{(f,+)}\right)=  +\frac{\pi }{L}
                                          \frac{1}{24}  \quad &
                                                                \text{if} \quad a=b 
\quad &  (f,+) \; {\rm BC} \, ,\\[1em]
- \frac{\pi }{L} \left(\frac{c}{24} - h_{(-,+)}\right) = +\frac{\pi }{L}
                                          \frac{23}{48} \quad & 
\text{if}\quad a>b \quad &  (-,+) \; {\rm BC} \, .
\end{array}\right.
\end{equation}
Since the three possible surface RG fixed points ($+$, $-$ or f) are
realized in configuration II, it
is interesting to study the RG flow between them. While we shall look at
this in greater detail in Sec.~V, we present here simple scaling
arguments for perturbations of the boundary condition around the free (f)
boundary fixed point. We thus start from the case $a=b$, which corresponds effectively to free (f) boundary
condition on the upper edge. The action $S$ on the strip can be formally decomposed as
\begin{equation}
S = S_{\text{bulk}}+ S_{\text{top edge}}+ S_{\text{bottom edge}} \, .
\end{equation}
To perturb this boundary
condition, one can add a field to the top edge,
\begin{equation}
S_{\text{top edge}}\to S_{\text{top edge}} +\lambda \int d x \;\phi(x),
\label{eq:pert_edge}
\end{equation}
where $\phi(x)$ is a local operator on the edge, with scaling
dimension $h_{\phi}$. For $b$ close to $a$, the coupling constant $\lambda$ must vary linearly with
$a-b$. One can see from the action (\ref{eq:pert_edge})
that $\lambda$ has scaling dimension $1-h_{\phi}$, so it must scale as
$\lambda \sim (a/b-1) (a+b)^{h_\phi-1}$.  Thus, there is a natural
length scale $\xi_c$ with
\begin{equation}
\label{eq:def_tau}
\xi_c(\tau) \sim |\tau|^{\frac{1}{h_{\phi}-1}}(a+b), \qquad \tau =
a/b-1 \, .
\end{equation}
At length scales smaller than $\xi_c$, the boundary condition is
effectively free, and at scales larger than $\xi_c$, it is effectively
fixed. We will refer to $\xi_c$ as the {\it crossover length}.  Here,
for the Ising universality class, we need to determine the perturbing
operator $\phi$ that generates an RG flow from free to fixed boundary
conditions and its scaling dimension $h_\phi$.  Clearly, the most natural
candidate for that operator is the spin operator itself, which breaks
$\mathbb{Z}_2$ symmetry. At the boundary, it has scaling
dimension $h_\phi = 1/2$ \cite{Cardy:1989xe}, so we obtain a crossover scale and
an associated critical exponent $\nu_c$, given by
\begin{equation}
\label{eq:scalingxi}
\xi_c(\tau) \sim |\tau|^{-\nu_c}(a+b), \quad \qquad \nu_c =2 \, .
\end{equation}
This length scale determines the scaling of the free energy which is
always a function of $L$ and of two dimensionless parameters, which we
can chose as $(a+b)/L$ and $\xi_c/L$.  We are interested in the regime
where $L, \xi_c \gg a,b$, such that
\begin{equation}
  \label{eq:II general}
  {\cal F}_{{\rm CFT}}(L,(a+b)/L,\xi_c/L) \;  \underset{L, \xi_c \gg a,b}{\sim} \;    {\cal F}_{{\rm CFT}}(L,0,\xi_c/L) ,
\end{equation}
leaving a function of two parameters only. It must take a
scaling form, in terms of a universal function $\vartheta$,
\begin{equation}
  \label{eq:3}
  \frac{{\cal F}_{{\rm CFT}}(L,0,\xi_c/L)}{W} = \frac{1}{L} \vartheta[L/\xi_c(\tau)] \, .
\end{equation}
In Sec.~V, we shall compute the free energy exactly, and 
recover the value of the critical exponent $\nu_c =2$. In addition, we will obtain
the scaling function $\vartheta$ explicitly.

\section{Determinants of real antisymmetric Block-Toeplitz 
matrices: the role of the Kitaev $\mathbb{Z}_2$ index}

In Sec.~V, we will see that the problem of calculating the partition
function for the strip with periodically alternating boundary
conditions requires the evaluation of the asymptotics of a large
block-Toeplitz determinant. This is a standard problem for which there
exist well-established results. Here, we briefly discuss some facts
about Toeplitz determinants. 

Interestingly, in addition to those well-established results, we need
to deal with one rather subtle case, involving real antisymmetric
block-Toeplitz matrices, for which the standard theorems (namely,
variants of the strong Szeg\"o limit theorem
\cite{widom1976asymptotic,bottcher2013analysis}) are not 
sufficient for our purposes. Here we shall simply state our main
  results that we need subsequently in this work, deferring all the
  details to a subsequent paper \cite{Basor:hb}.  A key role
  is played by Kitaev's pfaffian invariant \cite{kitaev2001unpaired},
  or $\mathbb{Z}_2$ invariant which distinguishes between two topologically
  distinct sets of real antisymmetric matrices.



\subsection{(Block-)Toeplitz matrices and the (common) strong Szeg\"o
  limit theorem}

Let $G_n$ be a {\it block-Toeplitz} matrix of size $n m   \times n m$, with $n^2$ blocks of size $m \times m$,
\begin{equation}
	\label{eq:Toep}
	G_n = \left( \begin{array}{c|c|c|c|c}
		g_{0} & g_1 & g_2  & \dots & g_{n-1} \\ \hline
		g_{-1} & g_0 & g_1  & \dots & g_{n-2} \\ \hline
		g_{-2} & g_{-1} & g_0  & \dots & g_{n-3} \\ \hline
		\vdots & \vdots & \vdots &  \ddots  \\ \hline
		g_{-n+1} & g_{-n+2} &  g_{-n+3} & & g_{0}
	\end{array} \right)  \, ,
\end{equation}
where each $g_p$ is an $m \times m$ complex matrix. The asymptotics of
$\det G_n$ is given by the strong Szeg\"o limit theorem
\cite{widom1976asymptotic}, which is a generalization of the strong
Szeg\"o limit theorem available for the scalar case ({\it i.e.}
$m=1$), also sometimes called the Szeg\"o-Widom theorem. Let us
briefly recall the statement of the strong Szeg\"o limit theorem. A
fundamental role in the theory of Toeplitz operators is played by the
Fourier transform of $g_p$, called the {\it symbol} in the mathematics
literature,
\begin{equation}
	\varphi(\theta) \, = \, \sum_{p \in \mathbb{Z}} e^{i p \theta} g_p , \qquad  \theta \in [0,2\pi] \simeq S^1.
\end{equation}
(In the literature, it is customary to introduce the symbol $\varphi$
first, and then view the sequence of matrices $G_n$'s as truncations
of the infinite-dimensional operator defined by $\varphi$.) When the
entries of $g_p$ decay sufficiently fast ({\it e.g.} exponentially)
with $p$, $\theta \mapsto \varphi(\theta)$ is a smooth function from
the circle $[0,2\pi] \simeq S^1$ to the space of $m \times m$ complex
matrices. One key assumption that is required is that
$\varphi(\theta)$ never vanishes (or equivalently, $\varphi(\theta)$
is invertible). If this holds, then one can define the winding number
$\mathcal{I} \in \mathbb{Z}$ of the map
\begin{equation}
	\label{eq:winding}
	\begin{array}{ccc} S^1 & \longrightarrow & \mathbb{C} \setminus \{ 0 \} \\
		\theta &\mapsto& \det \varphi(\theta) \, ,
	\end{array}
\end{equation}
which is also referred to as the {\it Fredholm index}. Indeed, it can
be shown that the Toeplitz operator with symbol $\varphi$ is Fredholm
iff $\det \varphi (\theta) \neq 0$ for all $\theta$, so that $\varphi$
is a (continuous) map from $S^1$ to ${\rm GL}(m, \mathbb{C})$, and
that the Fredholm index then labels the elements of the fundamental
group of
${\rm GL}(m, \mathbb{C}) \simeq {\rm U}(1) \times {\rm
  SL}(m,\mathbb{C})$.
Since ${\rm SL}(m,\mathbb{C})$ is contractible, this is nothing but
$\pi_1 ({\rm U}(1)) \simeq \mathbb{Z}$, and the index is precisely
what is counted by the winding number of Eq.~(\ref{eq:winding}) -- up
to a sign, depending on the convention for the orientation of the
winding -- see for instance
\cite{atiyah1969algebraic}. 

The content of the strong Szeg\"o limit theorem is the following (for
the original statement, see \cite{widom1976asymptotic}). If the
winding number $\mathcal{I}$ is zero, the leading asymptotics is given
by
\begin{equation}
	\label{eq:Szego}
	(\mathcal{I} = 0) \qquad \quad \frac{1}{n} \log \det G_n \; \underset{n\rightarrow \infty}{\longrightarrow} \; \int_0^{2\pi} \frac{d \theta }{2\pi} \log \det \varphi(\theta) .
\end{equation}
When $\mathcal{I} \neq 0$, this result is modified; references where
the case of non-zero index is discussed include
\cite{bottcher2013analysis,bottcher2006szego}. For scalar Toeplitz
matrices ({\it i.e.} $m=1$), this is the complete classification: the
Fredholm index decides whether Eq.~(\ref{eq:Szego}) applies ot
not. However, in the block-Toeplitz case ($m>1$), it remained somewhat
unclear whether references such as \cite{widom1976asymptotic} are
making additional assumptions to reach formula (\ref{eq:Szego});
different references on the strong Szeg\"o limit theorem for
block-Toeplitz matrices seem to be relying on different assumptions
(compare, for instance, the statements of the strong Szeg\"o limit
theorem in the block Toeplitz case in
\cite{widom1976asymptotic,bottcher2013analysis,deift2013toeplitz}). In
what follows, we will refer to Eq.~(\ref{eq:Szego}) as the {\it strong
  Szeg\"o limit formula} (as opposed to {\it theorem}), and discuss
whether or not it applies to the matrices that we will encounter later
on in this paper.

\subsection{An example involving a real antisymmetric matrix: does the
  strong Szeg\"o limit formula apply?}
\label{sec:2classD}

Let $u,v$ be real numbers, with $0<u <1$ and $0< v<1$. Consider the 
block-Toeplitz matrix (with $n^2$ blocks of size $2 \times 2$)
\begin{equation*}
	G_n = \left( \begin{array}{cc|cc|cc|c|cc}
		0 & u  & u v  & u^2 v & u^2 v^2 & u^3 v^2 & \dots & u^{n-1} v^{n-1} &   u^{n} v^{n-1}   \\ 
		-u & 0 & v  & u v & u v^2  & u^2 v^2 & \dots &  u^{n-1} v^n &  u^{n-1} v^{n-1}  \\ \hline
		- uv & - v & 0 & u  &   uv & u^2 v   & \dots & u^{n-2} v^{n-2} &   u^{n-1} v^{n-2}   \\ 
		-u^2 v &  - uv &  -u & 0 &  v & u v  &  \dots  & u^{n-2} v^{n-1} &  u^{n-2} v^{n-2}    \\ \hline
		- u^2 v^2 & - u v^2 & - u v  & -v &  0  & u  & \dots & u^{n-3} v^{n-3} &   u^{n-2} v^{n-3}    \\ 
		-u^3 v^2 &  - u^2v^2 &  -u^2 v & -u v &   -u & 0   & \dots  &  u^{n-3} v^{n-2} &  u^{n-3} v^{n-3}    \\ \hline
		\vdots & \vdots & \vdots &  \vdots   & \vdots & \vdots &   \ddots    &  \vdots & \vdots   \\ \hline
		-u^{n-1} v^{n-1} & - u^{n-1} v^n  &  -u^{n-2} v^{n-2} & - u^{n-2} v^{n-1}  &  -u^{n-3} v^{n-3} & - u^{n-3} v^{n-2}  &  \dots &  0 & u  \\
		-u^n v^{n-1}  &  -u^{n-1} v^{n-1}  & -u^{n-1} v^{n-2}  &  -u^{n-2} v^{n-2}  &  -u^{n-2} v^{n-3}  &  -u^{n-3} v^{n-3}  & \dots & -u & 0 
	\end{array} \right) \, .
\end{equation*}
One can check that, for any $n\geq 1$,
\begin{equation}
	\label{eq:exact_det_D2}
	\det G_n \,= \, u^{2n} \, .
\end{equation}
This is an exact result for any finite size, so there is of course no
need to use more advanced techniques to evaluate the asymptotics of
this determinant. However, it is instructive to see what the outcome
of the strong Szeg\"o limit formula  in that case is. First, notice
that the entries of $g_p$ are zero unless $p=-1,0,1$, so that the
assumption about the decay of the entries of the matrix $g_p$ is
trivially satisfied. In fact, the symbol is easily calculated, and it
is clearly a smooth function of $\theta$,
\begin{equation}
	\label{eq:exact_symbol_D2}
	\varphi(\theta) \, = \,  \frac{1}{\cos \theta - (uv + u^{-1}v^{-1})/2} \left( \begin{array}{cc}
		i \sin \theta  &   \frac{v+v^{-1}}{2}  - e^{-i \theta } \frac{u+u^{-1}}{2} \\
		-\frac{v+v^{-1}}{2}  + e^{i \theta } \frac{u+u^{-1}}{2} 	& i \sin \theta 
	\end{array} \right) .
\end{equation}
Second, one observes that, as long as $u \neq v$, the determinant of the symbol is non-zero,
\begin{equation*}
	\det \varphi(\theta) \, = \,  \frac{\cos \theta - (uv^{-1} + u^{-1}v)/2}{\cos \theta - (uv + u^{-1}v^{-1})/2}  .
\end{equation*}
Third, the winding number (or Fredholm index) is obviously
$\mathcal{I}=0$, because the determinant is real-valued and
non-zero. So all the basic assumptions of the strong Szeg\"o limit
formula, as stated above, are satisfied.  What result do we get when
we apply formula (\ref{eq:Szego})? The right hand side of Eq.~(\ref{eq:Szego})
is given by an integral which is readily evaluated,
\begin{equation}
	\label{eq:int_example}
	\int_0^{2\pi} \frac{d \theta }{2\pi} \log \left[  \frac{\cos \theta - (uv^{-1} + u^{-1}v)/2}{\cos \theta - (uv + u^{-1}v^{-1})/2} \right] \, = \, \oint_{|z|=1} \frac{dz}{2 \pi i} \frac{1}{z} \log \left[ \frac{(z-u^{-1}v )(z-u v^{-1})}{(z-u v)(z-u^{-1}v^{-1})} \right] \, = \, 2\, \log \left[ {\rm max}(u,v) \right]  \, .
\end{equation}
Hence, when $u > v$, the strong Szeg\"o limit formula yields
\begin{equation*}
	\frac{1}{n} \log \det G_n \; \underset{n\rightarrow \infty}{\longrightarrow} \; 2 \log u \, ,
\end{equation*}
in agreement with Eq.~(\ref{eq:exact_det_D2}). However, when $v>u$,
the strong Szeg\"o limit formula must be modified.

\subsection{A variant of the strong Szeg\"o limit formula}

In fact, real antisymmetric matrices require a modified version of the
strong Szeg\"o limit formula, which we state now, and illustrate with
the help of the example above. We need to introduce the Kitaev
$\mathbb{Z}_2$ invariant. Notice that, because $G_n$ is real
antisymmetric, the symbol has the following properties,
\begin{equation*}
	\varphi(\theta) \, = \, \varphi(\theta + 2\pi) \, = \, \varphi^*(-\theta) \, = \, -\varphi^\dagger(\theta) \, .
\end{equation*}
In particular, $\varphi(0)$ and $\varphi(\pi)$ are real and antisymmetric. One can then define the following number,
\begin{equation}
	\label{eq:Kitaev}
	\mathcal{I}_{2} \, = \, {\rm sign} \left[ {\rm Pf} \varphi(0) \; {\rm Pf} \varphi(\pi) \right]  \; \in \, \{ 1, -1 \} \, ,
\end{equation}
which is called the Kitaev pfaffian (or $\mathbb{Z}_2$) invariant. Pf
denotes the Pfaffian of the matrix $\varphi$. For a proof
that this number is a topological invariant, in the sense that it depends
continuously on the entries of the symbol as long as
${\rm det} \,\varphi \neq 0$, see
\cite{kitaev2001unpaired}. \vspace{0.5cm}

In this paper, we will use the following two formulas for the
asymptotics of the determinant of real antisymmetric block-Toeplitz
matrices (with $m \times m$ blocks, for $m$ even) that have no
additional symmetries. If $\mathcal{I}_{2} = +1$, then
\begin{subequations}
\begin{equation}
	\label{eq:Szego_as}
	(\mathcal{I}_{2} = +1) \qquad \quad \frac{1}{n} \log \det G_n \; \underset{n \rightarrow \infty}{\longrightarrow} \; \int_0^{2\pi} \frac{d \theta }{2\pi} \log \det \varphi(\theta) ,
\end{equation}
which is identical to Eq.~(\ref{eq:Szego}). If $\mathcal{I}_{2} = -1$,
then we have instead the modified formula
\begin{equation}
	\label{eq:Szego_as2}
	(\mathcal{I}_{2} = -1) \qquad \quad \frac{1}{n}  \log \left[  \det ( G_n ) /  \det \left( \int \frac{d\theta}{2\pi}  e^{i n \theta}  \varphi^{-1} (\theta) \right)  \right] \; \underset{n\rightarrow \infty}{\longrightarrow} \; \int_0^{2\pi} \frac{d \theta }{2\pi} \log \det \varphi(\theta) .
\end{equation}
\end{subequations}
Notice that this resolves the problem with the strong Szeg\"o
  limit formula in our above example. Indeed, in that example, one
has $\mathcal{I}_{2} = {\rm sign } \left[ u-v\right]$, such that when
$v>u$, one needs to use the modified formula (\ref{eq:Szego_as2}). The
entries of the Fourier transform of $\varphi^{-1}(\theta)$ all decay
as $(u v^{-1})^n$, because the decay rate is set by the complex zero
of $\det \varphi (\theta)$ that is closest to the real axis. As a
consequence, we see that
\begin{equation*}
	 \det \left( \int \frac{d\theta}{2\pi} e^{i  n \theta} \varphi^{-1}(\theta) \right)  \, \sim \, (u v^{-1})^{2 n} \, ,
\end{equation*}
up to a multiplicative constant. Hence, Eqs.~(\ref{eq:Szego_as2}) and
(\ref{eq:int_example}) give
$ \frac{1}{n} \log \left[\det (G_n)/ (u v^{-1})^{2n} \right] \,
\underset{n \rightarrow + \infty}{\rightarrow} \, 2 \log v$,
in agreement with the exact determinant of Eq.~(\ref{eq:exact_det_D2}).

\section{Ising universality class with periodically alternating 
boundary conditions: exact results}

In this section we come back to the Ising system that we defind and studied in
Sec.~\ref{Ising_uc} in some limiting cases. Here, we make use of the
theory of Toeplitz determinants presented in the previous section, to
obtain exact expressions for the Casimir free energy of the Ising
strip configurations. 
It turns out that, because we are considering the Ising universality
class with boundary conditions changing from $+$ to $-$, it is
possible to evaluate the $N$-point correlator of Eq.~(\ref{eq:bcc_correl})
explicitly. Indeed, in this case the BCC operator is the chiral part of the
energy operator in the Ising field theory, which can by identified
with the fermion $\psi$ in the Majorana fermion formulation. This field has
scaling dimension $h_{(+,-)} = h_{\psi} = 1/2$. The crucial
observation is that $\psi$ is a free field, so its $N$-points
correlator can be computed exactly from Wick's theorem. It has the
form of the Pfaffian of an $2N \times 2N$ antisymmetric matrix,
\begin{equation}
 \left< \psi(w_1) \dots \psi(w_{2N})  \right>\, = \, {\rm Pf}\, G \,,  \qquad \quad  G_{ij} \, = \,  \left< \psi(w_i) \psi(w_j) \right> .
\end{equation}
Using the above expression in Eq.~(\ref{eq:bcc_correl}), and making
use of the fact that the Ising model has central charge $c=1/2$, the
universal contribution to the free energy takes the form
\begin{equation}
	\mathcal{F}_{\rm CFT} = -\frac{\pi }{48} \frac{W}{L}  -\frac{1}{2}\log \left| \text{det} \;G \right|  .
	\label{Ising_gen_express}
\end{equation}
The calculation of the free energy, and therefore of the critical
Casimir force, is hence reduced to the evaluation of the determinant
of the $2N\times 2N$ matrix $G$ in the limit $N\to\infty$.  All
information on the interaction is contained in the two-point function
$\left< \psi(w_i) \psi(w_j) \right>$, which depends on the
homogeneous boundary conditions of the strip before we insert
BCC operators. Without loss of generality, we assume that, initially
({\it i.e.}~before the insertion of BCC operators), the boundary
conditions are fixed to $+$ on both sides of the strip. The two point
function is then
\begin{equation}
	 \left< \psi(z_i) \psi(z_j) \right> \, = \,
         \frac{1}{\frac{2L}{\pi} \, 
\sinh \frac{\pi (z_i-z_j)}{2L} }.
\end{equation}
As we shall see below, for both configurations of boundary conditions
(see Eqs. (\ref{eq:confI})-(\ref{eq:confII})), $G$ takes the form of a
real antisymmetric block-Toeplitz matrix. This allows us to evaluate
the large-$N$ asymptotics of the determinant, using the results
given in the previous section.

\subsection{Configuration I: periodically alternating boundary
  conditions on both sides}

\begin{figure}[h]
\begin{center}
\begin{tikzpicture}[scale=0.8]
	\filldraw[gray] (-8,0) rectangle (6,2);
	\draw[thick] (-8,0) -- (6,0);
	\draw[thick] (-8,2) -- (6,2);
	\filldraw (-6,0) circle (1mm);
	\filldraw (-4.5,0) circle (1mm);
	\filldraw (-3,0) circle (1mm);
	\filldraw (-1.5,0) circle (1mm);
	\filldraw (0,0) circle (1mm);
	\filldraw (1.5,0) circle (1mm);
	\filldraw (3,0) circle (1mm);
	\filldraw (4.5,0) circle (1mm);
	\filldraw (-7,2) circle (1mm);
	\filldraw (-5.5,2) circle (1mm);
	\filldraw (-4,2) circle (1mm);
	\filldraw (-2.5,2) circle (1mm);
	\filldraw (-1,2) circle (1mm);
	\filldraw (0.5,2) circle (1mm);
	\filldraw (2,2) circle (1mm);
	\filldraw (3.5,2) circle (1mm);
	\draw (-6.5,-0.2) node[below]{$\delta$};
	\draw[<->] (-6.9,-0.2) -- (-6.1,-0.2);		
	\draw (-8,1) node[left]{$L$};
	\draw[dashed] (-7,0) -- (-7,2);
	\draw[<->] (-6.9,2.2) -- (-5.6,2.2);
	\draw (-6.3,2.2) node[above]{$a$};
	\draw[<->] (-5.4,2.2) -- (-4.1,2.2);
	\draw (-4.8,2.2) node[above]{$a$};
	\draw[<->] (-5.9,-0.2) -- (-4.6,-0.2);	
	\draw (-5.2,-0.2) node[below]{$a$};
        \draw[<->] (-4.4,-0.2) -- (-3.1,-0.2);	
	\draw (-3.7,-0.2) node[below]{$a$};
	\draw (2.8,1.7) node{$+$};
	\draw (1.3,1.7) node{$-$};
	\draw (-0.2,1.7) node{$+$};
	\draw (-1.7,1.7) node{$-$};
	\draw (-3.2,1.7) node{$+$};
	\draw (-4.7,1.7) node{$-$};
	\draw (-6.2,1.7) node{$+$};
	\draw (3.8,0.3) node{$+$};
	\draw (2.3,0.3) node{$-$};
	\draw (0.8,0.3) node{$+$};
	\draw (-0.7,0.3) node{$-$};
	\draw (-2.2,0.3) node{$+$};
	\draw (-3.7,0.3) node{$-$};
	\draw (-5.2,0.3) node{$+$};
\end{tikzpicture}
\end{center}
\caption{Configuration I: Ising model on an infinite strip with
  periodically alternating boundary conditions on both sides. The
  number of up and down spins is equal on each side. $\delta$ denotes
  a lateral shift between the boundaries.}
	\label{fig:confI}
\end{figure}
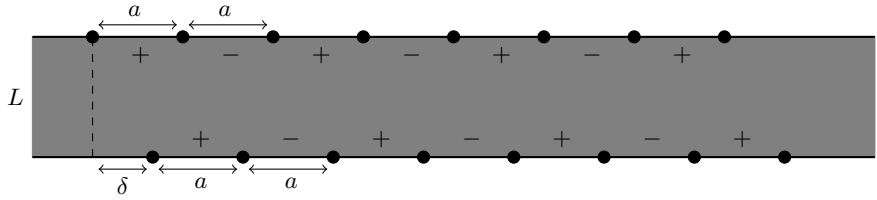

Consider configuration I, illustrated in Fig.~\ref{fig:confI} and
introduced in Sec.~\ref{Ising_uc}. Without loss of generality, we can
assume that $0 \leq \delta \leq a$. We are interested in the free
energy per unit length,
calculated in the limit $N \rightarrow +\infty$,
\begin{equation}
  \label{eq:partition2}
  \frac{{\cal F}_{\text{CFT}}}{W} = -\frac{\pi}{48 L} - \lim_{N \to\infty}
  \frac{1}{ 2N a} \log \left| \det G_{\text{I}} \right| \, ,
\end{equation}
where we used that the total length of the strip is $W = N a$.  The
matrix $G_{\text{I}}$ is block-Toeplitz, see
Eq.~(\ref{eq:Toep}), with the $2 \times 2$ blocks given by
\begin{subequations}
\begin{equation}
	(p\neq 0) \qquad   \left( \begin{array}{cc}
		\left< \psi(w_{2j}) \psi(w_{2j+2p}) \right> & \left< \psi(w_{2j}) \psi(w_{2j+2p+1}) \right> \\
		\left< \psi(w_{2j+1}) \psi(w_{2j+2p}) \right> & \left< \psi(w_{2j+1}) \psi(w_{2j+2p+1}) \right>
	\end{array} \right) = \left( \begin{array}{cc}
		\frac{1}{  \frac{2L}{\pi} \sinh \frac{\pi a p}{2L} } & \frac{1}{  \frac{2L}{\pi} \sinh (\frac{\pi (a p+\delta)}{2L}  - i \frac{\pi}{2} ) }  \\
		 \frac{1}{  \frac{2L}{\pi} \sinh (\frac{\pi (a p-\delta)}{2L}  + i \frac{\pi}{2} ) }  & \frac{1}{  \frac{2L}{\pi} \sinh \frac{\pi a p}{2L} }
	\end{array} \right) ,
\end{equation}
\begin{equation}
	(p=0) \qquad \left( \begin{array}{cc}
		0 & \left< \psi(w_{2j}) \psi(w_{2j+1}) \right> \\
		\left< \psi(w_{2j+1}) \psi(w_{2j}) \right> &0 
	\end{array} \right) = \left( \begin{array}{cc}
		0 & \frac{1}{  \frac{2L}{\pi} \sinh (\frac{\pi \delta}{2L}  - i \frac{\pi}{2} ) }  \\
		 \frac{1}{  \frac{2L}{\pi} \sinh (-\frac{\pi \delta}{2L}  + i \frac{\pi}{2} ) }  & 0
	\end{array} \right) .
\end{equation}
\end{subequations}
Notice that this matrix has complex entries. However, it is easily
transformed into a real matrix by conjugating it with the diagonal
matrix $\text{diag}(1,i,1,i,\cdots)$, which leaves the determinant
unchanged. Hence, the problem is reduced to evaluating the large $N$
asymptotics of the determinant of the real antisymmetric matrix $G_I$, defined as
\begin{subequations}
\begin{equation}
\label{eq:matrix_G_i}
	G_I = \left( \begin{array}{c|c|c|c|c}
		g_{0} & g_1 & g_2  & \dots & g_{N-1} \\ \hline
		g_{-1} & g_0 & g_1  & \dots & g_{N-2} \\ \hline
		g_{-2} & g_{-1} & g_0  & \dots & g_{N-3} \\ \hline
		\vdots & \vdots & \vdots &  \ddots  \\ \hline
		g_{-N+1} & g_{-N+2} &  g_{-N+3} & & g_{0}
	\end{array} \right) ,
\end{equation}
\begin{equation}
	(p\neq 0) \qquad  g_p = \left( \begin{array}{cc}
		\frac{1}{  \frac{2L}{\pi} \sinh \frac{\pi a p}{2L} } & \frac{1}{ \frac{2L}{\pi} \cosh (\frac{\pi (a p+\delta)}{2L}  ) }  \\
		 \frac{1}{  \frac{2L}{\pi} \cosh (\frac{\pi (a p-\delta)}{2L} ) }  & \frac{1}{  \frac{2L}{\pi} \sinh \frac{\pi a p}{2L} }
	\end{array} \right) ;  \qquad {\rm and} \quad   	 \qquad  g_0  = \left( \begin{array}{cc}
		0 & \frac{1}{ \frac{2L}{\pi} \cosh (\frac{\pi \delta}{2L}  ) }  \\
		 \frac{1}{ \frac{2L}{\pi} \cosh (\frac{\pi \delta}{2L}  ) }  & 0
	\end{array} \right) .
\end{equation}
\end{subequations}
The symbol ({\it i.e.}~the Fourier transform of $g_p$) is given by
\begin{equation}
\label{eq: gammasI}
 \varphi_{I}(\theta) = \frac{\pi}{L} \left(   \begin{array}{cc} i\; \gamma^I_1(\theta) &  (\gamma^I_2(\theta))^*  \\  \gamma^I_2(\theta) &  i\; \gamma^I_1 (\theta) \end{array}  \right), \qquad 
  \gamma^I_{1}(\theta)=\sum_{p =1}^{\infty} \frac{\sin (p \theta)}{\sinh (\frac{\pi  a p}{2 L})}\, ,\qquad  \gamma^I_{2}(\theta)=  \frac12 \sum_{p \in \mathbb{Z}} \frac{e^{ i p \theta}}{\cosh (\frac{\pi( p a - \delta)}{2 L})}\, . 
\end{equation}	
The coefficients $g_p$ decay exponentially with $p$, so the
symbol $\varphi_I(\theta)$ is an analytic function of
$\theta$. One observes that
\begin{equation}
	\det \varphi_I (\theta) \neq 0 \qquad {\rm if} \quad \delta \neq \frac{a}{2},
\end{equation}
while, if $\delta = \frac{a}{2}$, then $\det \varphi_I (\theta)$
vanishes at $\theta = \pi$. From now on, we assume
$\delta \neq \frac{a}{2}$. Then the map
$\theta \mapsto \det \varphi_I(\theta)$ has winding number
$\mathcal{I}=0$ [see Eq. (\ref{eq:winding})]. However, it is easy to
check that the Kitaev $\mathbb{Z}_2$ invariant is
\begin{equation}
	\mathcal{I}_2\,=\, \text{sign}[\gamma^I_{2}(0)\gamma^I_{2}(\pi)] \, = \, {\rm sign} \left[ a/2 - \delta \right] .
\end{equation}
Therefore, we need to distinguish two cases, according to the
discussion of Sec.~IV. When $\delta < a/2$, we can apply
Eq.~(\ref{eq:Szego_as}), while if $\delta > a/2$, we need to apply
Eq.~(\ref{eq:Szego_as2}) instead.  The exact expression for the free
energy is then
\begin{equation}
	\label{eq:exact_result_confI}
	\frac{\mathcal{F}_{\rm CFT}}{W} \, = \, \left\{  \begin{array}{lll} \displaystyle - \frac{\pi}{48 L} - \frac{1}{2 a} \int_0^{2\pi} \frac{d\theta}{2\pi} \log  \det \varphi_I(\theta)     &{\rm if}& \delta < \frac{a}{2}    \\   \displaystyle - \frac{\pi}{48 L} - \frac{1}{2 a} \left( \int_0^{2\pi} \frac{d\theta}{2\pi} \log  \det \varphi_I(\theta) \,+ \, \kappa \right)   &{\rm if}& \delta > \frac{a}{2}   ,\end{array} \right.
\end{equation}
where the additional term $\kappa$ of the modified Szeg\"o formula for
$\delta > a/2$ is
\begin{equation}
\label{Eq:kappa_I}
	\kappa \, = \, -\lim_{n \rightarrow +\infty} \frac{1}{n} \log  \det \left( \int \frac{d\theta}{2\pi} e^{i n \theta} \varphi_I^{-1}(\theta) \right)  .
\end{equation}

\subsubsection{Asymptotics for $L \ll a,\delta$}

For small separations, $L \ll a,\delta$, we can approximate
$\sinh \frac{\pi a p}{2L}$ by $\frac{1}{2} e^{\frac{\pi a p}{2L}}$,
and $\cosh \frac{\pi (p a-\delta)}{2L}$ by
$\frac{1}{2}e^{\frac{\pi (p a-\delta)}{2L}}$. Then the matrix is
equivalent to the one we treated in Sec.~\ref{sec:2classD}, see
Eq.~(\ref{eq:exact_symbol_D2}), with
$u = e^{- \frac{\pi \delta}{2 L}}$ and
$v= e^{- \frac{\pi (a-\delta)}{2L}}$. We see from
Eq.~(\ref{eq:exact_det_D2}) that the determinant is always given by
$u^{2N}= e^{- N\frac{\pi \delta}{L}}$. This leads to the expected
result of Eq.~(\ref{eq:smallLI}) for the free energy,
\begin{equation}
	\label{eq:confI_asympt_smallL}
\frac{\mathcal{F}_{\text{CFT}} }{W}= -\frac{\pi}{48 L}+ \frac{\pi}{2
  L} \frac{\delta}{a} + \ldots \, ,
\end{equation}
where we have ignored again subleading logarithmic
corrections. Interestingly, to leading order, the Casimir force is
attractive for $0<\delta<a/24$ and repulsive for $a>\delta>a/24$.

\subsubsection{Asymptotics for $L \gg a,\delta$}

\begin{figure}[h]
\begin{center}
\begin{tikzpicture}
\draw (0,0) node[left]{\includegraphics[scale=0.4]{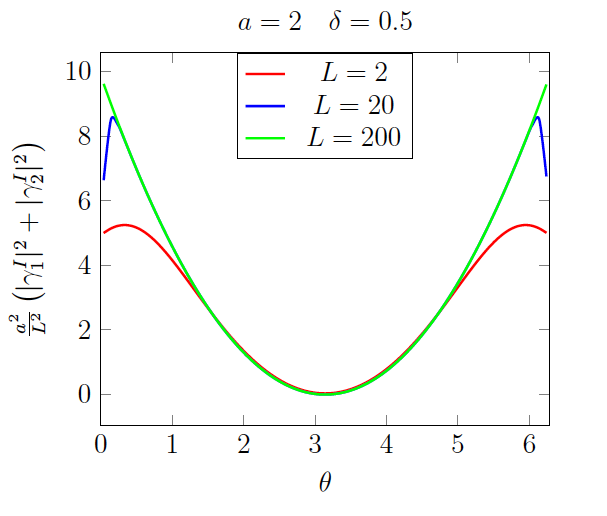}};
\end{tikzpicture}
\caption{This plot shows that, in the limit $L\to \infty$, the function
  $a^2/L^2 \text{det} \varphi_{I}(\theta)$ converges to a function
  $g(\theta)$ independent of $L$, see text for details.}
	\label{fig:check_detPhiI}
\end{center}
\end{figure}

Next, we turn to the regime where the width $L$ of the strip is much
larger than the modulation length $2a$ of the boundary conditions and
the lateral shift $\delta$ between the two boundaries. Let us start by
evaluating the integral of $\log {\rm det } \,\varphi_I$ from $0$ to
$2\pi$. The asymptotics of the functions $\gamma_{1}^{I}(\theta)$ and
$\gamma_{2}^{I}(\theta)$ can be obtained directly from the
Eqs.~(\ref{Eq:gamma1_largeL_th}), (\ref{Eq:hat_gamma}) and
(\ref{Eq:gamma3_largeL_th}) in the Appendix [with the substitutions
$\beta \rightarrow \pi a/(2L)$ and
$\alpha \rightarrow \pi \delta/(2L)$],
\begin{subequations}
\label{eq:asympt_gamma1_gamma2}
\begin{eqnarray}
\gamma_{1}^{I}(\theta) &\underset{L \gg a,\delta}{=}&  \frac{L}{a} \left[1-\frac{\theta}{\pi}+\tanh\left(\frac{L \theta}{a}\right)+\tanh\left( \frac{L (\theta-2\pi)}{a}\right)\right]+ O\left( 1\right)  \\[1.5em]
\gamma_{2}^{I}(\theta) &\underset{L \gg a,\delta}{=}&   \begin{cases}
    \frac{L}{a}\left[e^{i \theta \frac{\delta}{a}}\cosh\left(\frac{L \theta}{a}\right)^{-1}
\right] ,& \text{if }\; 0\leq\theta \leq \pi\\[1em]
   \frac{L}{a}\left[e^{i (\theta-2\pi) \frac{\delta}{a}}\cosh\left(\frac{L (\theta-2\pi)}{a}\right)^{-1}
\right] .              & \text{if} \;-\pi< \theta < 0
\end{cases} +O(1)\, .
\end{eqnarray}
\end{subequations}
One can check that
$\gamma_{1}^{I}(\theta)^2+|\gamma_{2}^{I}(\theta)|^2$ behaves as
$(L/a)^{2} g(\theta)$, where the function $g(\theta)$ does not depend
on $L$, as shown in Fig.~\ref{fig:check_detPhiI}.  Thus,
\begin{eqnarray*}
	\int_{0}^{2\pi} \frac{d\theta}{2\pi} \log | {\rm det} \, \varphi_I (\theta) |  & = & \log \left( (\pi / L)^2 \right) + \int_0^{2\pi} \frac{d\theta}{2\pi} \log [ \gamma_{1}^{I}(\theta)^2+|\gamma_{2}^{I}(\theta)|^2]  \\
	 & \underset{L \gg a,\delta}{=} & \log \left( (\pi / a)^2 \right) +   \int_0^{2\pi} \frac{d\theta}{2\pi} \log g(\theta)   
\end{eqnarray*}
which is a constant independent of $L$. If $\delta < a/2$, this is
sufficient to conclude that
$\lim_{N \rightarrow \infty} \frac{1}{N} {\rm det} G_I$ is a constant
independent of $L$. It is straightforward to verify from
Eq.~(\ref{eq:asympt_gamma1_gamma2}) that the
dominant term in the large $L$ limit of
$ \text{det} \varphi_{I}(\theta)$ does not depend on $\delta$,
i.e., the lateral Casimir force must decay faster than any power of $L$.

In the case $\delta > a/2$, however, we need to discuss the correcting
term in Eq.~(\ref{eq:Szego_as2}).  We have to evaluate the decay rate
$\kappa$ of the integral
$\int \frac{d\theta}{2\pi} e^{i \theta n} \varphi_{I}^{-1}(\theta)
\sim e^{-2n \kappa }$.
Since $\gamma^I_1(\theta)$ and $\gamma^I_2(\theta)$ are real analytic
in $\theta$, they must also be complex analytic in some finite strip
around the real axis in the complex plane. Assuming that
$\det \varphi_I (\theta)$ possesses at least one zero inside this
strip, we see from Cramer's formula that the singularity of
$\varphi^{-1}_I (\theta)$ that is closest to the real axis must be a
zero of $\det \varphi_I (\theta)$. The decay rate $\kappa$ is then
given by the imaginary part $\kappa =\Im(\theta_s)$ of the zero
$\theta_{s}$ of $\det \varphi_I (\theta)$ that is closest to the real
axis. One can see from
Eqs.~(\ref{eq:asympt_gamma1_gamma2}a)-(\ref{eq:asympt_gamma1_gamma2}b)
that, to leading order in $L$, this point is on the real axis,
$\theta_{s}=\pi$. Thus, $\kappa = 0$, and there is in fact no
correcting term to leading order in $L$. Hence, as in the case
$\delta < a/2$, for $\delta > a/2$ we find that
$\lim_{N\rightarrow \infty} \frac{1}{N} {\rm det} G_I$ is some
constant independent of $L$ in the regime $L \gg a,\delta$.  We
conclude that, for $L \gg a,\delta$, we recover the 
result for the free energy that we obtained above from simple scaling arguments,
\begin{equation}
\frac{\mathcal{F}_{\text{CFT}} }{W}=-\frac{\pi}{48 L}+ O\left( L^{-2}\right).
\end{equation}
As discussed in Sec.~\ref{Ising_uc}, this result is consistent with
the RG picture that the boundary conditions flow towards the same
homogeneous boundary conditions on both sides of the strip,
yielding an attractive normal force for all $\delta$. Hence, if
  $a>\delta>a/24$ then there must be a change from an attractive to a
  repulsive force when the separation $L$ is decreased, leading to a
  stable minimum of the free energy (at fixed $\delta$). This we shall
  see explicitly in the next section.

\subsubsection{Arbitrary separations $L$}

\begin{figure}[t]
	\begin{center}
	\includegraphics[width=0.6\textwidth]{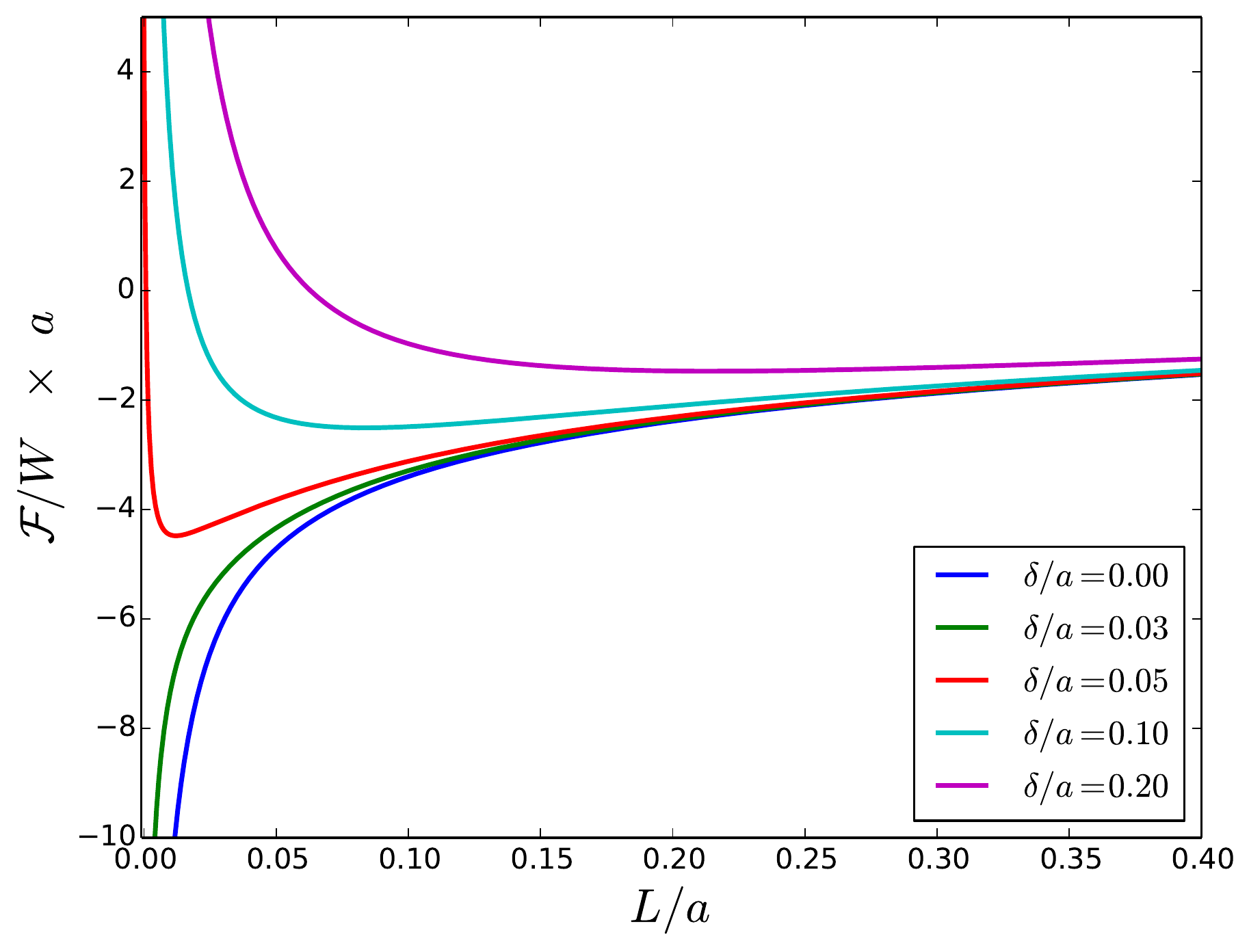}
        \caption{\label{fig:confI_general_L_2} Free energy per unit
          length (measured in units of $a$), as given by the exact
          formula of Eq.~(\ref{eq:exact_result_confI}), as a function
          of $L$, for different values of $\delta/a$. Note that the
          free energy shown here contains ($L$-independent) surface
          self-energies and hence does not vanish for
          $L\to\infty$. The self-energies depend on the boundary
          conditions itself but not the lateral shift $\delta$.}
	\end{center}
\end{figure}

\begin{figure}[t]
	\begin{center}
		(i) \includegraphics[width=0.32\textwidth]{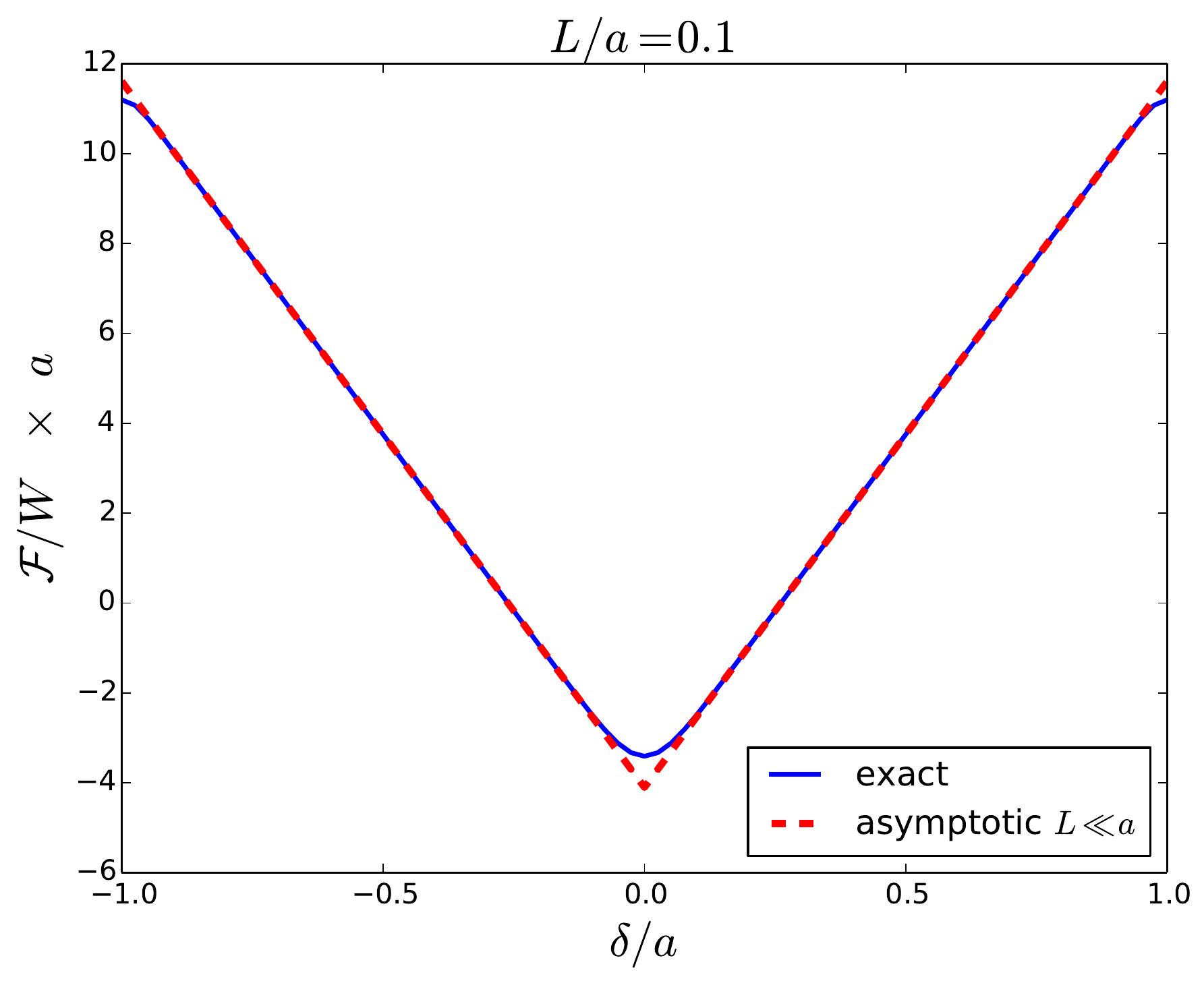}
		\includegraphics[width=0.32\textwidth]{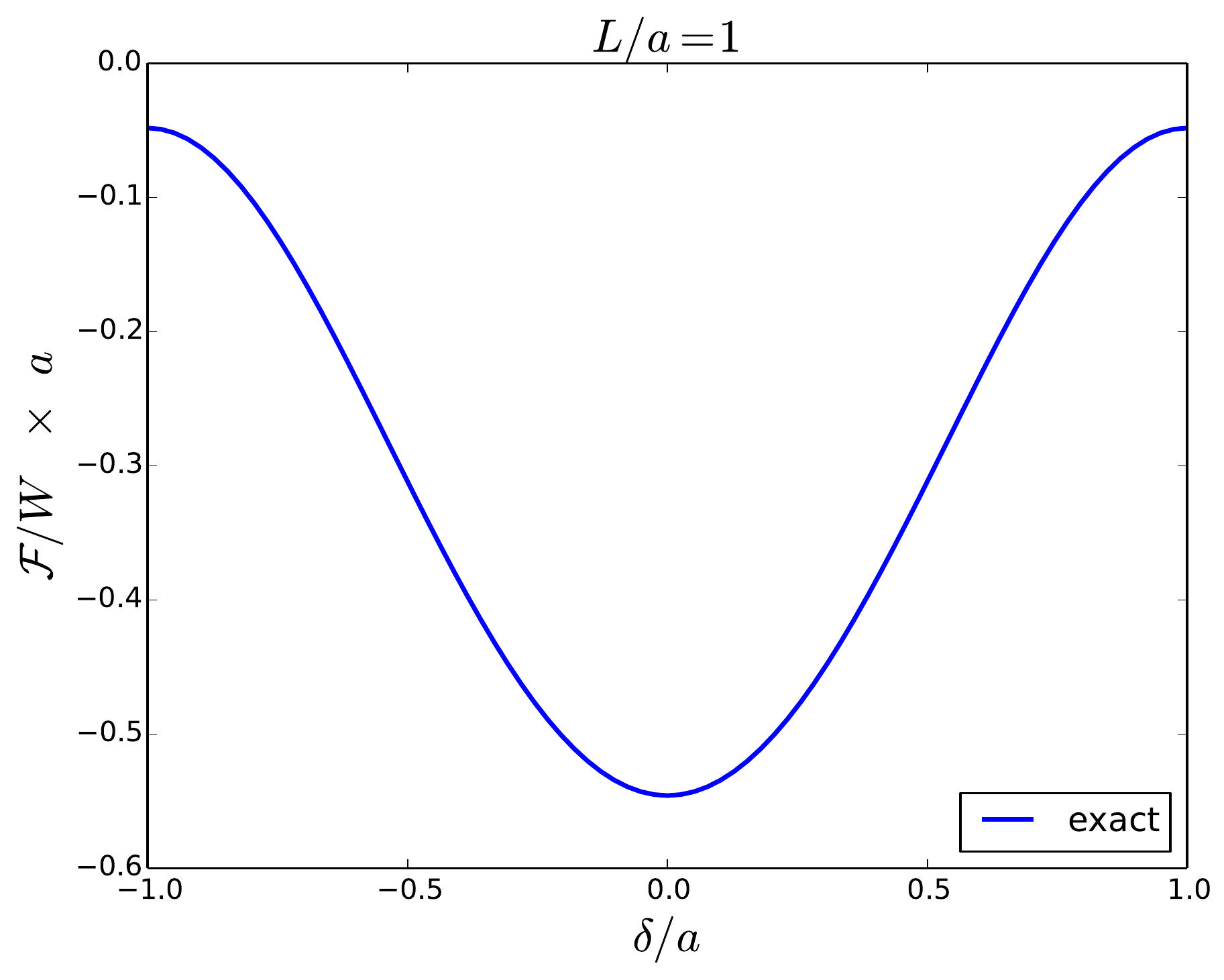}
		\includegraphics[width=0.32\textwidth]{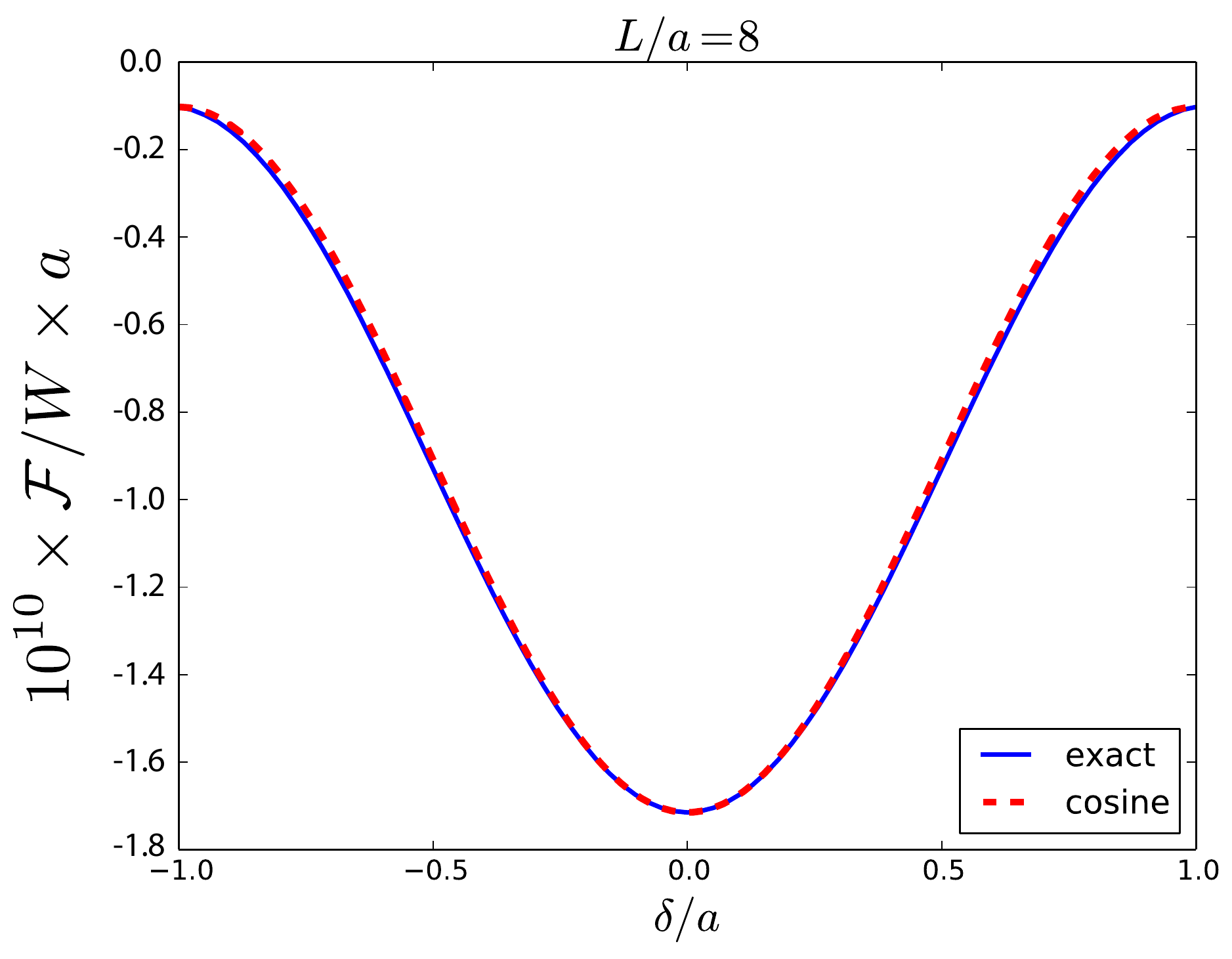}
		(ii) \includegraphics[width=0.5\textwidth]{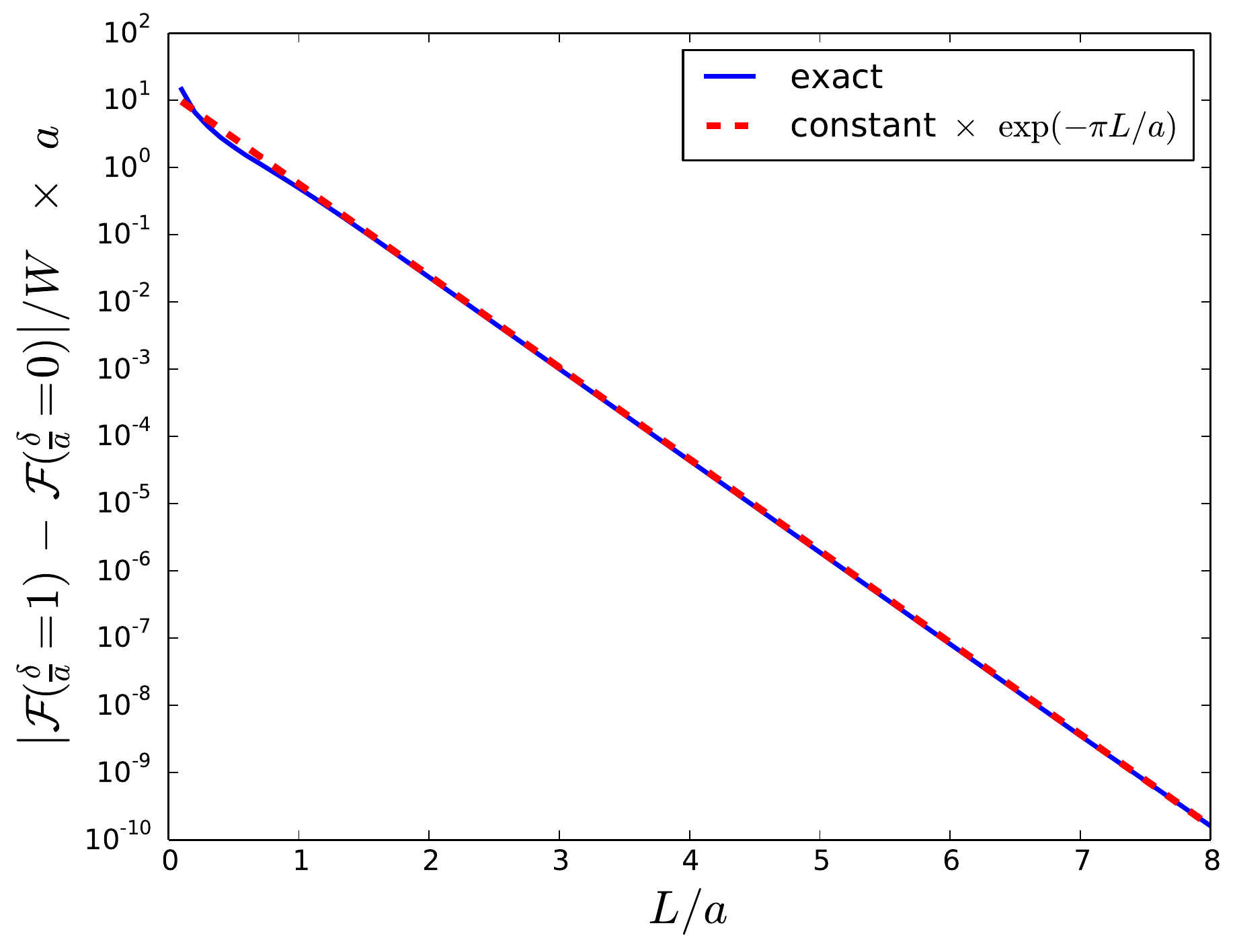}
                \caption{\label{fig:confI_general_L} (i) Free energy
                  per unit length (measured in units of $a$), as given
                  by the exact formula of
                  Eq.~(\ref{eq:exact_result_confI}), as a function of
                  $\delta$, for different values of $L$. The left
                  curve ($L/a = 0.1$) is well approximated by the
                  asymptotic formula of
                  Eq.~(\ref{eq:confI_asympt_smallL}) valid for
                  $L \ll a$, while the right curve ($L/a = 8$) is well
                  approximated by $\cos(\pi \frac{\delta}{a} )$ times
                  an $L$-dependent amplitude. For the latter case, we
                  have rescaled the free energy and subtracted a
                  constant offset. (ii) The 
                  amplitude of the periodic modulation, defined as
                  $\frac{1}{W} | \mathcal{F}(\frac{\delta}{a}=1) -
                  \mathcal{F}(\frac{\delta}{a}=0) |$,
                  decays as $e^{- \pi \frac{L}{a}}$ at large $L$.}
	\end{center}
\end{figure}

To study the Casimir interaction at arbitrary separations, we have
evaluated the free energy from the (modified) Szeg\"o formula of
Eq.~\eqref{eq:exact_result_confI} by numerical summation of the series
and integration over $\theta$. In addition, we have obtained the free
energy directly from a numerical computation of the determinant of
truncated versions of the infinite matrix $G_I$ [see
Eq.~\eqref{eq:matrix_G_i}] and extrapolation of the results to
$N\to\infty$. Both methods yield coincident results. The free energy
per unit length is shown in Fig.~\ref{fig:confI_general_L_2} as
function of the rescaled separation $L/a$ for different values of the
lateral shift $\delta$. As expected from the limiting behavior of the
free energy and small and large separations $L$, the free energy shows
a minimum if $a>\delta > a/24$. This minimum is most pronounced for
$\delta$ slightly larger than $a/24$, and for larger $\delta$ becomes
more shallow and displaced to larger $L/a$. Hence, be tuning the
lateral shift of the boundaries, different equilibrium positions of
the boundaries can be achieved.

In order to study the lateral variation of the Casimir interation, we
have also computed the free energy per unit length for three
different fixed separations ($L/a=0.1, \, 1.0, \, 8.0$) as function of
the lateral shift over one period ($-a \le \delta \le a$), see
Fig.~\ref{fig:confI_general_L}(i). At short separations, $L/a=0.1$, we
recover the approximation of Eq.~\eqref{eq:confI_asympt_smallL} with
good accuracy. Towards intermediate distances, $L/a=1$, the dependence
on $\delta$ becomes more cosine like and finally at sufficiently large
separations $L/a=8$, we find convergence to a perfect
$\cos(\pi \delta/a)$ dependence on $\delta$. This simple cosine
dependence of the Casimir force has been observed before for QED
Casimir forces between geometrically structured surfaces at large
separations, irrespective of the precise form of the surface's
periodic pattern \cite{Buscher:2005db}. Similarly, for the Ising strip
(and other CFT models) we expect for more complicated but periodic
spin boundary conditions a convergence to such a simple cosine
dependence at large separations. This follows from the fact that the
free energy must be a periodic function of $\delta$ and hence can be
decomposed into a discrete Fourier series. As we shall see in a
moment, the amplitudes of the harmonics decay exponentially with
$2\pi L/\lambda$ where $\lambda = 2a/m$ is the wavelength of the
harmonic function of order $m$. Hence, higher harmonics with $m>1$ are
more strongly suppressed at large $L$, leaving only a cosine
dependence from $m=1$.  The dependence of the free energy on $\delta$
yields a lateral Casimir force
$F_\text{lat} = -\partial \mathcal{F}/\partial \delta$ that is
determined by the curves in Fig.~\ref{fig:confI_general_L}(i).

Fig.~\ref{fig:confI_general_L}(ii) shows the dependence of the
amplitude of the modulation of the free energy with $\delta$, and
hence the decay of the lateral force $F_\text{lat}$ with $L$. For
sufficiently large $L/a$ the amplitude and hence the lateral force
decays exponentially as $\sim e^{-\pi L/a}$ which is in agreement with
the decay expected for the lowest harmonic with $m=1$.

\subsection{Ising model with periodically alternating boundary
  conditions on one side}

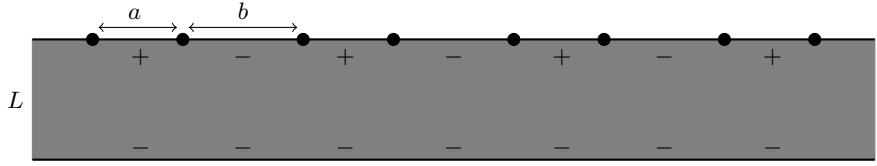
\begin{figure}[ht]
\begin{center}
\begin{tikzpicture}[scale=0.8]
	\filldraw[gray] (-8,0) rectangle (6,2);
	\draw[thick] (-8,0) -- (6,0);
	\draw[thick] (-8,2) -- (6,2);
	\filldraw (-7,2) circle (1mm);
	\filldraw (-5.5,2) circle (1mm);
	\filldraw (-3.5,2) circle (1mm);
	\filldraw (-2,2) circle (1mm);
	\filldraw (0,2) circle (1mm);
	\filldraw (1.5,2) circle (1mm);
	\filldraw (3.5,2) circle (1mm);
	\filldraw (5,2) circle (1mm);
	\draw (-8,1) node[left]{$L$};
	\draw[<->] (-6.9,2.2) -- (-5.6,2.2);
	\draw (-6.3,2.2) node[above]{$a$};	
	\draw[<->] (-5.4,2.2) -- (-3.6,2.2);
	\draw (-4.5,2.2) node[above]{$b$};
	\draw (4.3,1.7) node{$+$};
	\draw (2.5,1.7) node{$-$};
	\draw (0.8,1.7) node{$+$};
	\draw (-1,1.7) node{$-$};
	\draw (-2.8,1.7) node{$+$};
	\draw (-4.5,1.7) node{$-$};
	\draw (-6.2,1.7) node{$+$};
	\draw (-6.2,0.2) node{$-$};
	\draw (-4.5,0.2) node{$-$};
	\draw (-2.8,0.2) node{$-$};
	\draw (-1,0.2) node{$-$};
	\draw (0.8,0.2) node{$-$};
	\draw (2.5,0.2) node{$-$};
	\draw (4.3,0.2) node{$-$};
\end{tikzpicture}
\end{center}
\caption{Configuration II: Ising model on an infinite strip with
  periodically alternating boundary conditions on one side, and
  homogeneous boundary conditions on the other side. On the
  upper boundary, the number of up spins (over length $a$) and down
  spins (over length $b$) can be different.}
	\label{fig:confII}
\end{figure}

Now we focus on the configuration II (see Sec.~\ref{Ising_uc}), and
we show in detail how to recover the results of Eqs.~(\ref{eq:II small L}) and
(\ref{eq: II big L}) from the analysis of the asymptotics of a
block-Toeplitz determinant. Notice that the total length of the strip is
now $W = N (a+b)$ in Eq.~(\ref{Ising_gen_express}), and the free energy
per unit length is then, for $N \rightarrow +\infty$,
\begin{equation}
  \label{eq:partition3}
  \frac{{\cal F}_{\text{CFT}}}{W} = -\frac{\pi}{48 L} - \lim_{N
    \to\infty}  \frac{1}{ 2N (a+b)} \log \left| \det G_{\text{II}} \right| \, .
\end{equation}
 The block-Toeplitz matrix $G_{\text{II}}$
 whose determinant we need to evaluate is defined by the following blocks,
\begin{subequations}
	\label{eq:block_gp_II}
\begin{equation}
	(p\neq 0) \qquad  g_p = \left( \begin{array}{cc}
		\left< \psi(w_{2j}) \psi(w_{2j+2p}) \right> & \left< \psi(w_{2j}) \psi(w_{2j+2p+1}) \right> \\
		\left< \psi(w_{2j+1}) \psi(w_{2j+2p}) \right> & \left< \psi(w_{2j+1}) \psi(w_{2j+2p+1}) \right>
	\end{array} \right) = \left( \begin{array}{cc}
		\frac{1}{  \frac{2L}{\pi} \sinh \frac{\pi (a+b) p}{2L} } & \frac{1}{  \frac{2L}{\pi} \sinh \frac{\pi (a+b)p +\pi a}{2L}  }  \\
		 \frac{1}{  \frac{2L}{\pi} \sinh \frac{\pi (a+b) p-\pi a}{2L} }  & \frac{1}{  \frac{2L}{\pi} \sinh \frac{\pi (a+b) p}{2L} }
	\end{array} \right) ,
\end{equation}
\begin{equation}
	 g_0 = \left( \begin{array}{cc}
		0 & \left< \psi(w_{2j}) \psi(w_{2j+1}) \right> \\
		\left< \psi(w_{2j+1}) \psi(w_{2j}) \right> &0 
	\end{array} \right) = \left( \begin{array}{cc}
		0 & \frac{1}{  \frac{2L}{\pi} \sinh \frac{\pi a}{2L}  }  \\
		 -\frac{1}{ \frac{2L}{\pi} \sinh \frac{\pi a}{2L} }  & 0
	\end{array} \right) .
\end{equation}
\end{subequations}

A brief account of our results for this configuration has been given
in Ref. \cite{Dubail:2016uo}. Below we shall give further details on
the derivation of the results.

\subsubsection{Asymptotics for small separations $L \ll a,b$}

In the regime of a narrow strip, $L \ll a,b$, we can approximate
$\sinh \frac{\pi (a+b) p}{2L}$ by
$\frac{1}{2} e^{\frac{\pi (a+b) p}{2L}}$, and write similar
expressions for the other entries of the blocks in
Eq.~\eqref{eq:block_gp_II}. Again, this case is exactly equivalent to
the one we treated in Sec.~\ref{sec:2classD} [see
Eq.~(\ref{eq:exact_symbol_D2})] now with
$u = e^{- \frac{\pi a}{2 L}}$ and $v= e^{- \frac{\pi b}{2L}}$. We saw
in Eq.~(\ref{eq:exact_det_D2}) that the determinant is always
$u^{2n}= e^{- n\frac{\pi a}{L}}$. Thus, we recover the
result of Eq.~(\ref{eq:II small L}),
\begin{equation}
\frac{\mathcal{F}_{\text{CFT}} }{W}= -\frac{\pi}{48 L}+ \frac{\pi}{2
  L} \frac{a}{a+b} 
+ \ldots \, ,
\end{equation}
where we have ignored again subleading logarithmic corrections.

\subsubsection{Asymptotics for large separations $L \gg a,b$}

The symbol of the block-Toeplitz matrix of Eq.~(\ref{eq:block_gp_II}) is
\begin{equation}
	\label{eq: gammasII}
\varphi_{II}(\theta) = \frac{\pi}{L} \left(   \begin{array}{cc} i \gamma^{II}_1(\theta) &  -(\gamma^{II}_2(\theta))^*  \\   \gamma^{II}_2(\theta) &  i \gamma^{II}_1 (\theta) \end{array}  \right) , \qquad \gamma^{II}_{1}(\theta)=\sum_{p =1}^\infty \frac{\sin (p \theta)}{\sinh \frac{\pi p (a+b)}{2 L}}\, ,\qquad  \gamma^{II}_{2}(\theta)=  \frac12 \sum_{p \in \mathbb{Z}} \frac{e^{ i p \theta}}{\sinh \frac{\pi( p (a+b) -  a)}{2 L}}\, .
\end{equation}
One can check that the index $\mathcal{I}_2$ is
\begin{equation}
\mathcal{I}_2= {\rm sign}[b-a] .
\end{equation}
When $L \gg a,b$, the asymptotics of the functions
$\gamma_{1}^{II}(\theta)$ and $\gamma_{2}^{II}(\theta)$ can be read
directly from the formulas (\ref{Eq:gamma1_largeL_th}),
(\ref{Eq:hat_gamma}) and (\ref{Eq:gamma2_largeL_th}) with the substitutions
$\beta \rightarrow \pi(a+b)/(2L)$ and $\alpha \rightarrow \pi a/(2L)$, yielding
\begin{subequations}
\begin{eqnarray}
  \label{eq:gamma1bigL}
    \gamma^{II}_1(\theta) &=& \frac{L}{a+b} \left\{
    \hat{\gamma}_{1}(\theta) + \tanh \left(\frac{L \theta}{a+b}\right)
+\tanh\left(\frac{L(\theta-2\pi)}{a+b}\right)\right\}\\
\label{eq:gamma2bigL}
\gamma^{II}_2(\theta) &=& \frac{L}{a+b} \left\{
    \hat{\gamma}_{2}(\theta,\tau) +i \left[\tanh \left(\frac{L \theta}{a+b}\right)
+\tanh\left(\frac{L(\theta-2\pi)}{a+b}\right)\right]\right\}
\end{eqnarray}
\end{subequations}
with
\begin{eqnarray}
  \hat{\gamma}_1(\theta) &=& 1-\frac{\theta}{\pi} \, , \nonumber \\
  \hat{\gamma}_2(\theta,\tau) &=& \frac{1}{\pi} \left[ 
-\frac{\tau+2}{\tau+1} + e^{i \theta} (\tau+2)
{}_2F_1\left(1,\frac{1}{\tau+2};\frac{\tau+3}{\tau+2};e^{i \theta}\right) - e^{-i \theta} \frac{\tau+2}{2\tau+3} 
{}_2F_1\left(1,\frac{2\tau+3}{\tau+2};\frac{3\tau+5}{\tau+2};e^{-i \theta}\right)\right]. \quad 
  \end{eqnarray}
  We recall that the parameter $\tau$ was defined as $\tau = a/b - 1$
  in Eq.~(\ref{eq:def_tau}). In the following we treat the cases $a<b$
  and $a>b$ separately since they differ in the index $\mathcal{I}_2$
  and hence require the application of different versions of the
  Szeg\"o limit formula.

  \noindent {\bf The case $a< b$ ($\mathcal{I}_2 = +1$).} Using the
  expressions (\ref{eq:gamma1bigL}) and (\ref{eq:gamma2bigL}), for $a<b$
  (or $\tau < 0$), Eq.~(\ref{eq:Szego}) yields the free energy density
  per unit length
\begin{equation}
  \label{Eq:Free_energy_II_largeL}
 \frac{ {\cal F}_{\text{CFT}}}{W} = -\frac{\pi}{48 L} -\frac{1}{4\pi(a+b)} \int_0^{2\pi} d\theta\; \log\left\{  1 + \Gamma(\theta,\tau) 
\left[ \tanh{\left(\frac{\theta L}{a+b}\right)} +\tanh{\left(\frac{(\theta-2\pi)L}{a+b}\right)} \right] \right\}  \,  ,
\end{equation}
where
\begin{equation}
\label{eq:def_Gamma}
\Gamma(\theta,\tau)=\frac{2\hat\gamma_1(\theta)+i(\hat\gamma_2(\theta,\tau)-\hat\gamma^*_2(\theta,\tau))}
{|\hat\gamma_1(\theta)|^2-|\hat\gamma_2(\theta,\tau)|^2}.
\end{equation}
To obtain Eq.~(\ref{Eq:Free_energy_II_largeL}) we have subtracted an
a term that is $L$-independent and hence does not contribute to the Casimir
force. Notice also that the $\log L$ term originating from the global
factor $\pi/L$ of $\varphi_{II}(\theta)$, see Eq.~(\ref{eq: gammasII}), is
cancelled by another one coming from the scaling with $L$ of
$\gamma^{II}_{1}(\theta)$ and $\gamma^{II}_{2}(\theta)$, see
Eqs.~(\ref{eq:gamma1bigL}) and (\ref{eq:gamma2bigL}). We note
  that the rescaled free energy
$L \mathcal{F}_{\rm CFT}$ depends only on $\tau$ and $(a+b)/L$.

\begin{figure}
\includegraphics[scale=0.4]{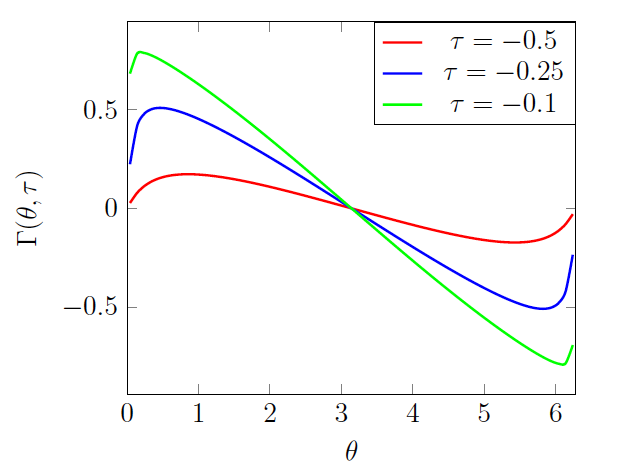}
\caption{Graph of the function $\Gamma(\theta,\tau)$, defined in
  Eq.~(\ref{eq:def_Gamma}), for different values of $\tau$.}
\label{Fig:plotGamma}
\end{figure}

The integrand in Eq.~(\ref{Eq:Free_energy_II_largeL}) is exponentially
localized around $\theta=0, 2\pi$ over a range $\sim
(a+b)/L$.
Therefore, it is important to consider the scaling behavior of the
function $\Gamma(\theta, \tau)$ around $\theta = 0$ and
$\theta= 2\pi$. This function is shown in Fig.~(\ref{Fig:plotGamma})
for different values of $\tau$.  Expanding the numerator and the
denominator of Eq.~(\ref{eq:def_Gamma}) around $\theta=0$ and $\tau=0$, the function $\Gamma(\theta,\tau)$ has the expansion
\begin{equation}
\Gamma(\theta,\tau) = \frac{\frac{2}{\pi}\theta}{\frac{\pi^2}{16}\tau^2+\frac{2}{\pi}\theta} + O\left(\theta^2,\tau \theta, \tau^3 \right),
\end{equation}
that implies the following scaling behavior when $\tau \to 0 $,
\begin{equation}
\label{Eq:Gamma_asy}
\Gamma\left(\zeta | \tau |^x,\tau \right) \underset{\tau \rightarrow 0}{=} \begin{cases}
    1 & \text{if $0 < x<2$},\\
     \frac{\zeta}{\zeta+\pi^3/32} & \text{if $x=2$}, \\
    0 & \text{if $x > 2$},
  \end{cases} 
\end{equation}
for any constant $\zeta$.  From this behavior of
$\Gamma(\theta, \tau)$ it is evident that the relevant length scale to
be compared to $L$ is $\xi_c(\tau) \sim (a+b) |\tau|^{-2}$ introduced in
Eq.~(\ref{eq:def_tau}) and Eq.~(\ref{eq:scalingxi}).  It is
  crucial to realize that the two limits $L/(a+b) \to \infty$ and
  $\tau \to 0$ do not commute. Hence, we need to distinguish three
  different scaling regimes for $\tau \to 0$, depending on the value
  of $L/\xi_c$ that we shall identify with $1/\zeta$:

\begin{enumerate}[(i)]

\item The regime $L \gg \xi_c$ or $L/(a+b) \sim \tau^{-x}$, $x>2$ (fixed
  spin BCs): The behavior of the integral in
  Eq.~(\ref{Eq:Free_energy_II_largeL}) depends on the limits of
  $\Gamma(\theta,\tau)$ when $\theta\to 0^+$ and $\theta\to 2\pi^-$,
  which are both zero so that the integral vanishes. The free energy per
  unit length is then to leading order $\mathcal{F}_{\rm CFT}/W = -\frac{\pi}{48 L}$,
  which corresponds to like fixed spin boundary conditions on both
  sides, as argued in Eq.~(\ref{eq: II big L}).

\item The regime $L \ll \xi_c$ or $L/(a+b) \sim \tau^{-x}$, $0<x<2$ (free-fixed
  spin BCs): The integral in Eq.~(\ref{Eq:Free_energy_II_largeL}) can be
  treated as follows. We consider first the part of integration from
  $0$ to $\pi$. We split the integration into two intervals,
  $0\leq \theta\leq C \tau^x$ and $C \tau^x\leq \theta\leq \pi$, for
  some fixed constant $C>0$. After a change of variables,
  $\theta \to \tilde{\theta}= L\theta/(a+b)$, the integral can be
  approximated by
\begin{eqnarray*}
&& -\frac{1}{4\pi(a+b)} \int_0^{2\pi} d\theta\; \log\left\{  1 + \Gamma(\theta,\tau) 
\left[ \tanh{\left(\frac{\theta L}{a+b}\right)} +\tanh{\left(\frac{(\theta-2\pi)L}{a+b}\right)} \right] \right\} \\
&& \qquad \simeq  -\frac{1}{4\pi L} \int_{C L\tau^x/(a+b))}^{\infty}  \log \left\{ 1+ \Gamma\left(\frac{(a+b)\tilde{\theta}}{L}, \tau\right)\left[ \tanh{\tilde{\theta}}-1    \right]\right\}\;d\tilde{\theta} \\
&& \qquad \qquad  -\frac{1}{4\pi L} \int_{0}^{C L\tau^x/(a+b))} \log \left\{ 1+ \Gamma\left(\frac{(a+b)\tilde{\theta}}{L}, \tau\right)\left[ \tanh{\tilde{\theta}}-1 \right]\right\}\;d\tilde{\theta}.
\end{eqnarray*}
Because of the factor $\tanh \tilde{\theta}-1$, the first integral is
localized around $C L\tau^x/(a+b)$ while the second term is of order
$C L\tau^x/(a+b))$, since the integrand is bounded.  Replacing
$\Gamma$ by $1$, see Eq.~(\ref{Eq:Gamma_asy}), and taking the limit
$C\to 0$, we see that the second integral vanishes, while the first
one becomes
\begin{equation*}
-\frac{1}{4\pi L}\int_{0}^{\infty} \; d\tilde{\theta} \log\left[ \tanh{(\tilde{\theta})}\right]=  \frac{\pi}{32 L}.
\end{equation*}
The same contribution comes from the integration from $\pi$ to $2\pi$
in Eq.~ (\ref{Eq:Free_energy_II_largeL}). Hence, the free energy
$\mathcal{F}_{\rm CFT}/W = - \frac{\pi}{48 L} + \frac{\pi}{16 L} =
\frac{\pi}{24 L}$,
as expected from Eq.~(\ref{eq: II big L}) for a strip with free spin
BCs on one and fixed spin BCs on the other side.

\item The scaling regime $L\sim \xi_c$ or $L/(a+b) \sim \tau^{-2}$ (flow from free spin BCs to fixed spin BCs):
  this is probably the most interesting case since it yields the scaling function
  that describes the flow from free to fixed boundary conditions. To
  evaluate the integral in Eq.~(\ref{Eq:Free_energy_II_largeL}), one
  proceeds again by splitting the integral into two parts, exactly as
  in the previous case. Replacing $\Gamma(\zeta\tau^2,\tau)$ by
  $\zeta/(\zeta+\pi^2/32)$, see Eq.~(\ref{Eq:Gamma_asy}), we find
  the following expression for the free energy per unit length,
 \begin{eqnarray}
\nonumber \frac{\mathcal{F}_{\text{CFT}}}{W} &=&
                                                 -\frac{\pi}{48}\frac{1}{L} - \frac{1}{L}\frac{1}{2\pi}\int_{0}^{\infty}\; d \tilde{\theta} \log\left\{1+ \frac{\zeta}{\zeta+\pi^2/32}(\tanh{(\tilde{\theta})}-1)\right\} \\
 &=&   \frac{1}{4\pi L}\text{Li}_2\left(  \frac{2 \zeta}{\zeta+\pi^3 /32}-1\right),
\end{eqnarray}
where $\text{Li}_2(x)$ is the polylogarithm function.  Hence,
identifying the scaling variable $\zeta$ with $\xi_c/L$, we conclude
that the exponent $\nu_c$ defined in Eq.~\eqref{eq:scalingxi} is
indeed $2$ and the free energy takes the form
\begin{equation}
\frac{\mathcal{F}_{\text{CFT}}}{W} =\frac{ \vartheta_{-}(\zeta)}{L},
\end{equation}
where the universal scaling function $\vartheta_{-}(\zeta)$ for $a<b$ is given by
 \begin{equation}
\label{eq:vartheta-}
 \vartheta_{-}(\zeta)=\frac{1}{4\pi}\text{Li}_2\left(
    \frac{2 \zeta}{\zeta+\pi^3 /32}-1\right) \, .
 \end{equation} 
\end{enumerate}

\begin{figure}
\begin{tikzpicture}
\draw (0,0) node[left]{\includegraphics[scale=0.25]{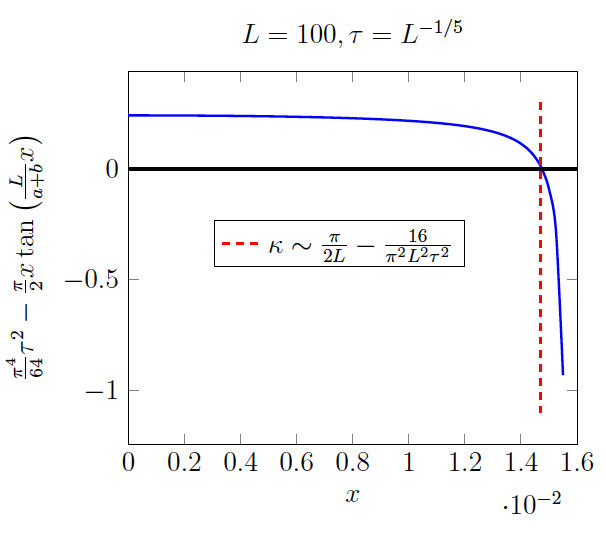}};
\draw (3,0) node{\includegraphics[scale=0.25]{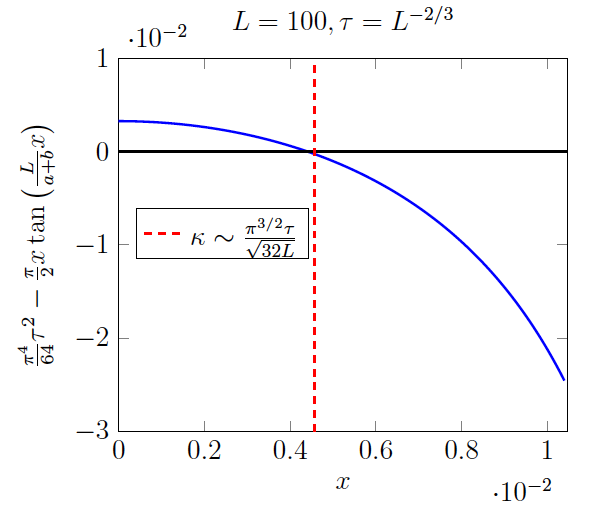}};
\draw (9,0) node{\includegraphics[scale=0.25]{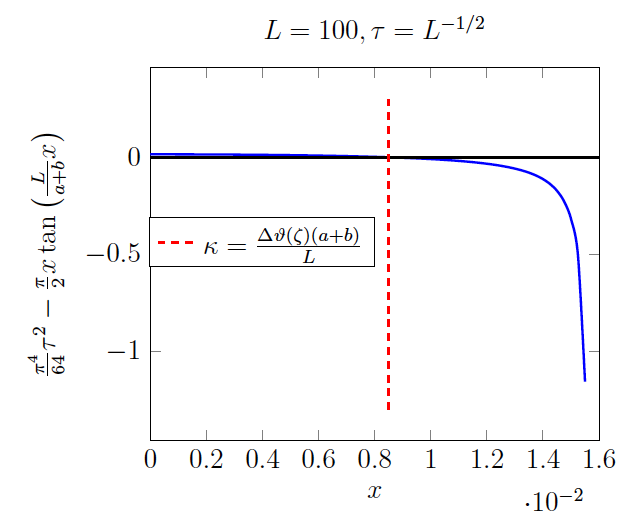}};
\end{tikzpicture}
\caption{\label{fig:3regimes} The panels show, from the left to the
  right, the solution of the Eq.~\eqref{Eq:kappaeqII} in the three
  regimes (i), (ii) and (iii) discussed in the text. In particular, the
  values of the exponent $x$, $L/(a+b)\sim \tau^{-x}$ have been set to
  $x=(5,3/2,2)$ for the regime (i), (ii) and (iii) respectively. The
  parameters $a$ and $b$ have been chosen such that $a+b=1$.}
\end{figure}

\noindent {\bf The case $a>b$ ($\mathcal{I}_2 = -1$).} In this case
the large $N$ asymptotic of the Toeplitz determinant is given by
Eq.~\eqref{eq:Szego_as2}. Note that the contribution to the free energy coming from the term $\int_{0}^{\pi} \frac{d \theta }{2\pi} \log \det \varphi_{II}(\theta)\; $ is symmetric in $a$ and $b$ and is still given by the Eq.~\eqref{Eq:Free_energy_II_largeL}. Analogously to the case of  configuration $I$ with $\delta >a/2$, see Eq.~\eqref{eq:exact_result_confI} and Eq.~\eqref{Eq:kappa_I},  we have to add to Eq.~\eqref{Eq:Free_energy_II_largeL} the decay rate $\kappa$:
\begin{equation}
\kappa = -\lim_{n\to \infty} \frac{1}{n}\log  \det \int_{0}^{2\pi} \frac{d \theta }{2\pi} e^{i n
  \theta} \varphi^{-1}_{II}(\theta) \, . 
\end{equation}
The free energy per unit length is therefore
\begin{equation}
\label{Eq:freeenergyIIagb}
\frac{\mathcal{F}_{\text{CFT}}}{W} = \text{r.h.s. of
  Eq.~\eqref{Eq:Free_energy_II_largeL}} + \frac{\kappa}{(a+b)} \, .
\end{equation} 
We recall that $\kappa$ is determined by the location of the nearest
pole $\theta_{s}$ to the real axis of the function
$[\varphi_{II}^{-1}(\theta)]_{11}$. The pole $\theta_{s}$ is obtained
from the zeros of the determinant, $\det \varphi_{II}(\theta_s)=0$. In
particular one has $\kappa = \Im(\theta_{s})$.  To determine $\kappa$,
we can use the expansions of Eqs.~(\ref{Eq:gamma1_largeL}) and
(\ref{Eq: gamma2_smalltheta}) with $\beta = \pi (a+b)/2L$ and
$\tilde{\epsilon}\sim \tau/2$, where $\tau$ is defined in
Eq.~\eqref{eq:def_tau}. We find that
\begin{equation}
\text{det}\; \varphi_{II}(\theta) =  \frac{\pi^4}{64} \tau^2 + 
 \frac{\pi}{2} \theta\; \tanh\left(\frac{L \theta}{a+b}\right)+O\left(\frac{\tau}{L}\right)
\end{equation}
The value of $\kappa$ is therefore given by the solution of the
non-linear equation
\begin{equation}
\label{Eq:kappaeqII}
\frac{\pi^4}{64} \tau^2 -\frac{\pi}{2} \kappa \tan\left( \frac{L
    \kappa}{a +b}\right) =0 \, .
\end{equation}
As shown in Fig.~(\ref{fig:3regimes}), the solutions of
Eq.~\eqref{Eq:kappaeqII} depend on the following three scaling regimes
for $\tau\to 0$:

\begin{enumerate}[(i)]
\item The regime $L \gg \xi_c$ or $L/(a+b) \sim\tau^{-x}$, $x>2$ (fixed -, fixed + spin  BCs). In this regime, the Eq.~\eqref{Eq:kappaeqII} yields
\begin{equation}
\kappa = \pi \frac{a}{L} +O(L^{2\frac{1-x}{x}})
\end{equation}
The free energy Eq.~\eqref{Eq:freeenergyIIagb} then becomes
$\mathcal{F}_{\text{CFT}}/W= -\frac{\pi}{48 L}+\frac{\kappa}{(a+b)} =
-\frac{\pi}{48 L}+\frac{\pi}{2 L}= +\frac{\pi}{L}\frac{23}{48}$
and we recover the $(-,+)$ BCs of Eq.~\eqref{eq: II big L}.

\item The regime $L\ll \xi_c$ or $L/(a+b) \sim \tau^{-x}$, $0<x<2$
  (free-fixed spin BCs). In this regime, the solution of
  Eq.~\eqref{Eq:kappaeqII} takes the form
\begin{equation}
\kappa  \sim  L^{-\frac{2+x}{2 x}} \, .
\end{equation}
The free energy Eq.~\eqref{Eq:freeenergyIIagb} is still given by
$\mathcal{F}_{\text{CFT}}/W= -\frac{\pi}{48 L}+\frac{\pi}{16 L}$ as
$\kappa$ gives a sub-leading contribution, $\kappa \sim
O(L^{-1})$.
Hence we recover thus the result Eq.\eqref{eq: II big L} for $(f,+)$
BCs.

\item The scaling regime $L\sim \xi_c$ or $L/(a+b) \sim \tau^{-2}$
  (flow from the free to fixed + spin BCs). We can write the
  non-linear equation for $\kappa$ in terms of the scaling variable
  $\zeta$ with $L/(a+b)=\zeta \tau^{-2}$
  in the compact form
\begin{equation}
 \Delta \vartheta(\zeta)  \tan\left[ \Delta \vartheta(\zeta) \right]
=\frac{\pi^3}{32} \zeta,
\end{equation}
where we defined
\begin{equation}
\frac{\Delta \vartheta}{L}  = \frac{\kappa}{(a+b)} \, .
\end{equation}
Using Eqs.~\eqref{eq:vartheta-} and \eqref{Eq:freeenergyIIagb}, the
universal scaling function $\vartheta_{+}(\zeta)$, defined by
\begin{equation}
\frac{\mathcal{F}_{\text{CFT}}}{W}= \frac{\vartheta_{+}(\zeta)}{L} 
\end{equation} 
is given by
\begin{equation}
\label{eq:vartheta+}
\vartheta_{+}(\zeta) =  \vartheta_{-}(\zeta)+\Delta \vartheta(\zeta)
\, .
\end{equation}

\end{enumerate}

\subsubsection{Arbitrary separations $L$}

\begin{figure}[t]
	\begin{center}
	\includegraphics[width=0.6\textwidth]{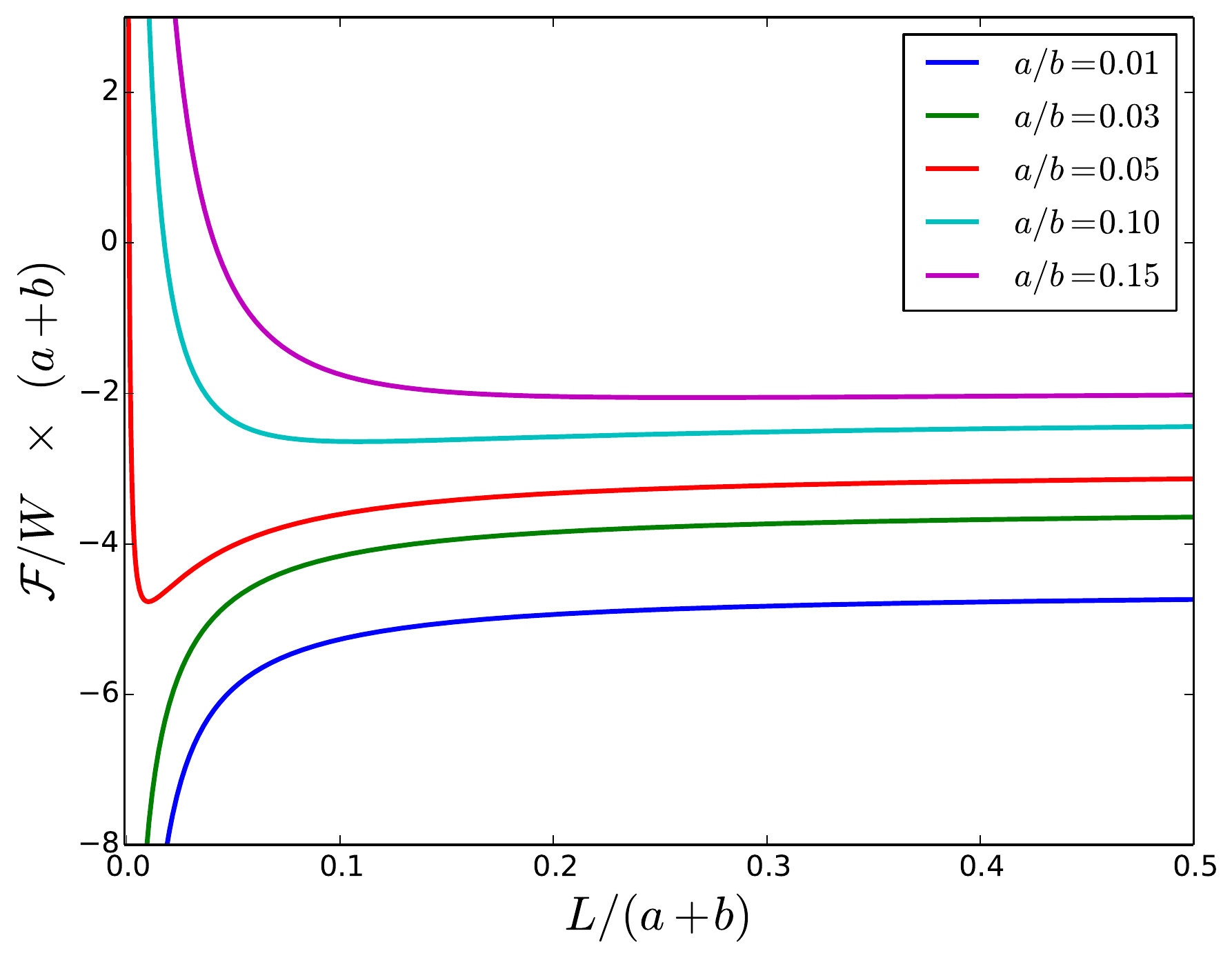}
        \caption{Free energy per unit length (measured in units of
          $a+b$), as given by the exact formula of
          Eq.~\eqref{eq:partition3} as function of $L$ for different
          values of $a/b$. Note that the free energy shown here
          contains ($L$-independent) surface self-energies and hence
          does not vanish for $L\to\infty$. The self-energies depend
          on the boundary conditions and hence 
          on $a/b$.}
\label{fig:force_sign_change_conf_II}
	\end{center}
\end{figure}

To study the free energy for arbitrary separations, and to determine
the position of the energy minimum for the cases with $b/23 < a < b$,
we have evaluated Eq.~\eqref{eq:partition3} numerically for different
ratios $a/b$, following the procedure outline above for configuration
I.The result is shown in Fig.~\ref{fig:force_sign_change_conf_II}. The
minimum in the free energy is most pronounced for $a/b$ slightly
larger than $1/23$. For increasing values of $a/b$ the minimum becomes
rather shallow.

\begin{figure}[t]
	\begin{center}
	\includegraphics[width=0.6\textwidth]{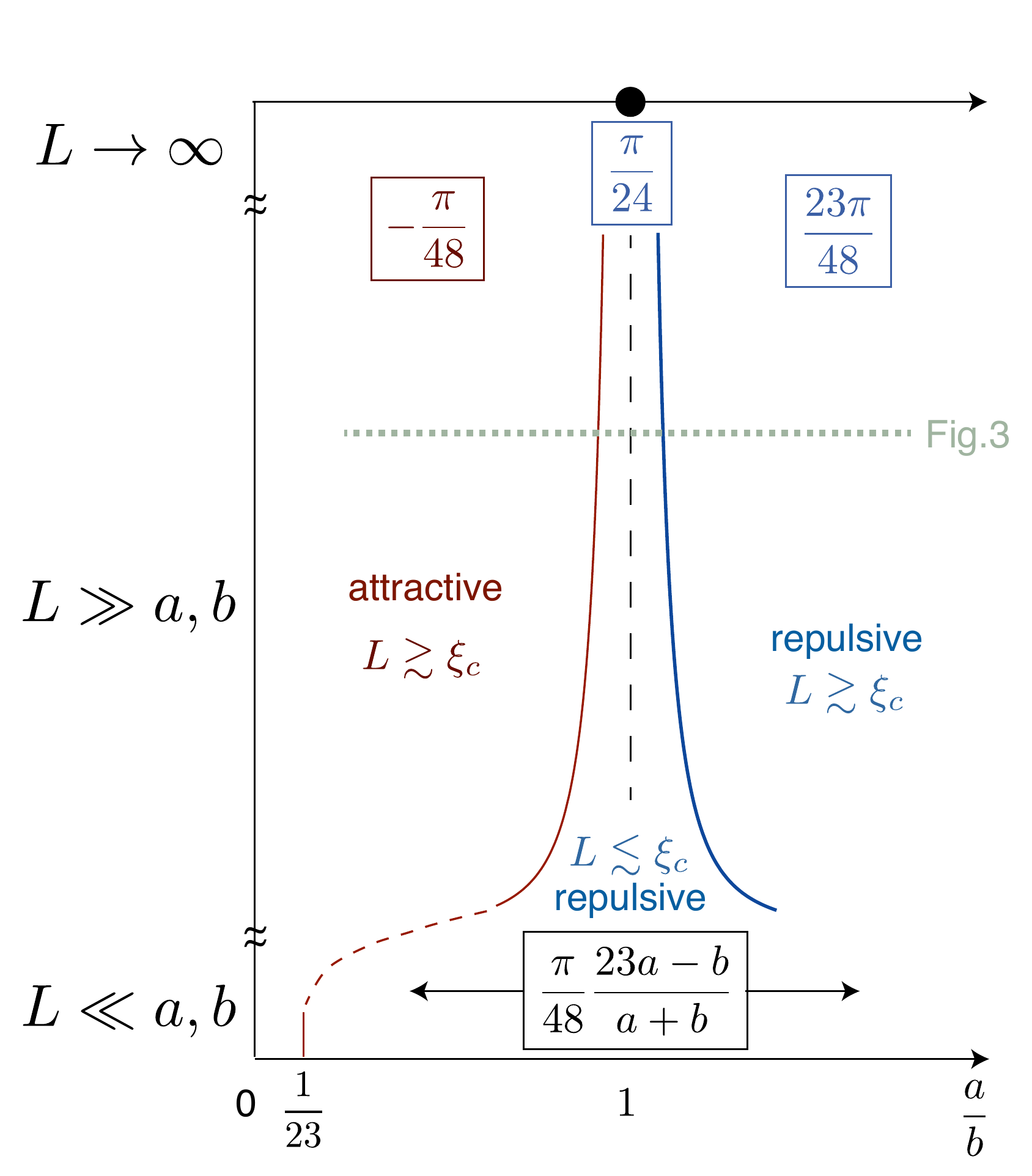}
       \caption{Schematic overview of critical Casimir force amplitudes as
  function of the strip width $L$ and the ratio $a/b$. For $L\gg a,b$ the
  solid curves represent the diverging crossover length $\xi_c$. The
  horizontal dashed line indicates the cut along which the force
  amplitude is plotted in Fig.~\ref{fig:scaling_fct}. Across the red
  curve the sign of the force changes whereas the blue curve indicates
only a change between two universal (repulsive) limits.}
\label{fig:scheme}
	\end{center}
\end{figure}

\begin{figure}[t]
	\begin{center}
	\includegraphics[width=0.6\textwidth]{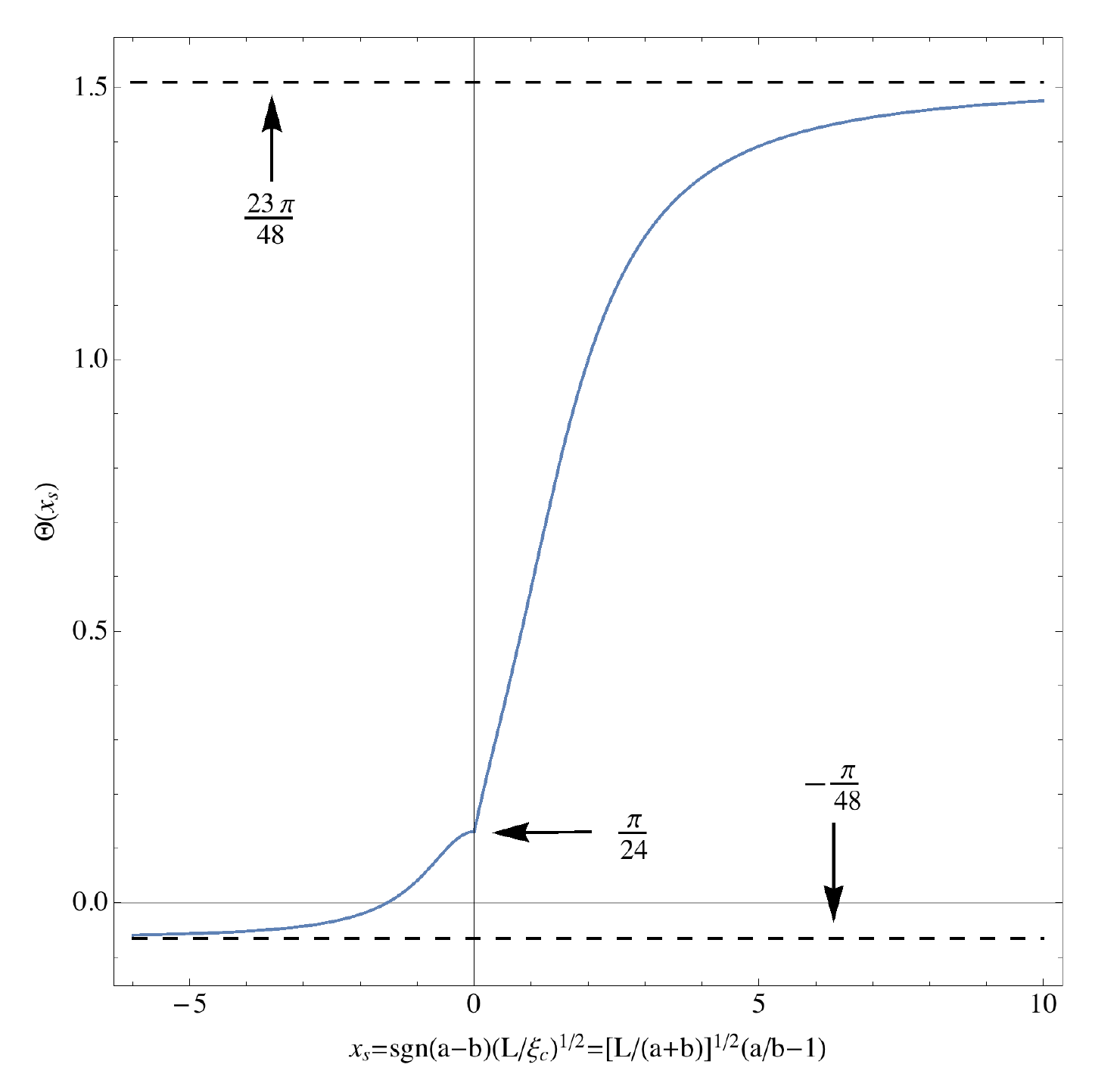}
        \caption{Universal scaling function $\Theta(x_s)$ for the critical
  Casimir force as function of the scaling variable $x_s\sim a-b$.}
\label{fig:scaling_fct}
	\end{center}
\end{figure}

Our overall findings for configuration II can be summarized by the
scheme of Fig.~\ref{fig:scheme}.  It shows the different scaling
regimes and the corresponding asymptotic amplitudes of the Casimir
force. At short distance $L\ll a,b$ the amplitude varies continuously
across the critical point at $a=b$, with a sign change at
$b/a=23$. For $L\gg a,b$ there exist three distinct regions: around
$a=b$ appears a region where $L \ll \xi_c$ where the force is
repulsive and approaches for asymptotic $L$ the universal amplitude
for fixed-free spin boundary conditions. For $a<b$, the force changes
sign from attractive to repulsive when $L$ approaches $\xi_c$,
corresponding to a stable point. For $a>b$, the force is always
repulsive but the amplitude crosses over from $\pi/24$ to $23\pi/48$
under an increase of $L$ beyond $\xi_c$.

The dependence of the force $F$ on $|a-b|$ at fixed $L\gg a,b$ (see
dashed horizontal line in Fig.~\ref{fig:scheme}) is determined by
$F = -\partial {\cal F}/\partial L = \Theta(x_s) L^{-2}$ with a
universal scaling function $\Theta$ of the scaling variable $x_s$ that
is defined on both sides of the critical point by
$x_s=\text{sign}(\tau)(L/\xi_c)^{1/2}\sim a-b$. This function is shown
in Fig.~\ref{fig:scaling_fct} where we used the results for
$\vartheta_\pm(\zeta)$ of Eqs.~\eqref{eq:vartheta-},
\eqref{eq:vartheta+}.  In the critical region $|x_s| \ll 1$, one has
the expansions
\begin{equation}
  \label{eq:16}
  \Theta(x_s)  = \left\{
\begin{array}{ll}
\frac{\pi}{24} - \frac{\pi^2}{64} x_s^2 + \ldots & \text{for } x_s < 0 \\[1em]
\frac{\pi}{24} + \frac{\pi^{3/2}}{8\sqrt{2}} x_s + \ldots & \text{for } x_s > 0 
\end{array}\right. \, ,
\end{equation}
whereas for $L$ outside the critical region, $|x_s| \gg 1$,
\begin{equation}
  \label{eq:17}
  \Theta(x_s)  = \left\{ \!
\begin{array}{ll}
-\frac{\pi}{48} + \frac{32\log 2}{\pi^4} \frac{1}{x_s^{2}} + \ldots & 
\text{for } x_s < 0 \\[1em]
\frac{23\pi}{48} - \frac{32(\pi^2-\log 2)}{\pi^4} \frac{1}{x_s^{2}} + 
\ldots & \text{for } x_s > 0 
\end{array}\right.  .
\end{equation}
We see that $\Theta(x_s)$ is not analytic around $x_s=0$ and hence
constitutes a singular part of the free energy density. This resembles
the singular nature of scaling functions describing the bulk
transition at $T=T_c$.

\section{Conclusion}

We have demonstrated that boundary CFT is a powerfull tool to study
critical Casimir forces in two dimensional systems with inhomogeneous
boundary conditions. We have studied explicitly a Ising strip at its
critical temperature with periodically varying boundary conditions
along the edges of the strip. Depending on the relative number and
position of the fixed spins at the edges, we have observed a number of
interesting phenomena: 

\begin{enumerate}[(i)]

\item At short separation between the edges, their Casimir interaction
  is determined by additivity, i.e., by a superposition of the
  interaction for like and unlike fixed spin boundary conditions.

\item If an edge has identical numbers of fixed up and down spins, at
  large separations it contributes to the Casimir interaction
  effectively as an edge with homogeneous free spin boundary conditions.

\item When the fixed spin boundary conditions break $\mathbb{Z}_2$
  symmetry (edge with unequal number of up and down spins) the
  effective boundary condition determining the Casimir force depends
  on the edge separation. The crossover in the Casimir interaction
  between the different boundary conditions is described by a
  universal scaling function.

\item The normal Casimir force between the edges can be attractive or
  repulsive. There are cases where the force changes sign from
  attractive to repulsive when the edge separation is reduced, leading
  to a stable equilibrium position. 

\item For the configurations studied here, a stable position (with
  respect to the normal separation $L$) exists for configuration I if
  the lateral shift $\delta$ obeys the condition $a/24 < |\delta| <a$
  (modulo periodicity) and for configuration II if the modulation
  length obey the condition $b/23 < a < b$.

\item When both edges have periodically modulated fixed spin boundary
  conditions, there acts a lateral Casimir force between them. This
  force decays exponentially with the edge separation over a length
  scale that is given by the wavelength of the boundary conditions. At
  short separations, the lateral force follows from additivity,
  leading to a piecewise constant force. At large separations, the
  lateral force assumes a universal form that follows a simple cosine
  profile.

\end{enumerate}

The observed renormalization of modulated fixed spin boundary
conditions, in binary mixtures, allows for an experimental realization
of ordinary (free spin) boundary conditions. These boundary conditions
can be ``switched'' on and off by varying the distance $L$, or an
inhomogeneous surface field. The position of the minimum of the free
energy can be tuned either by varying the lateral shift between the
boundaries or by changing the relative number of up and down spins.
This provides external control parameters to define a preferred
equilibrium separation between the surfaces.  It is interesting to
explore these concepts in three dimensions for Ising and XY models,
and tri-critical points which have an even richer spectrum of possible
boundary conditions. Previous mean field and numerical studies of a 3D
Ising system confined between parallel boundaries with alternating
boundary conditions indicated indeed the possibility of a sign change
of the force \cite{Toldin:2013uk}.

\vspace{0.5cm}

{\bf Acknowledgements.} We are grateful to Davide Vodola, Jean-Marie
St\'ephan, Eddy Ardonne, Estelle Basor and Karol Kozlowski for very
helpful discussions and exchange about section IV. We also thank
Andrea Gambassi and Mehran Kardar for general discussions.  This
work was partially supported (JD) by the Conseil G\'en\'eral de
Lorraine and the Universit\'e de Lorraine and by the CNRS "D\'efi
Inphyniti". JD thanks Nordita, Stockholm, for hospitality during the
workshop "From quantum field theories to numerical methods", where
part of this work was done.

\appendix

\section{Asymptotics infinite sums}
We define the functions:
\begin{equation}
f_1(x) = \frac{\sin (x \theta)}{\sinh (\beta x)},\quad  f^{\pm}_2(x) = \frac{\exp (\pm i x \theta)}{\sinh (\beta x\mp\alpha),}\quad f^{\pm}_3(x) = \frac{\exp (\pm i x \theta)}{\cosh (\beta x\mp\alpha)}
\end{equation}
and the corresponding sums:
\begin{equation}
\label{Eq:definitions}
\gamma_{1}(\theta) = \sum_{n\in \mathbb{Z}^{+}}^{\infty} f_1(n),\quad \gamma_{2}(\theta) = \sum_{n \in \mathbb{Z}}^{\infty} f^{+}_2(n)\quad \text{and} \quad \gamma_{3}(\theta) = \sum_{n \in \mathbb{Z}}^{\infty} f^{+}_3(n).
\end{equation}
We are interested in the asymptotics of the above sums in the limit:
\begin{equation}
\beta\to 0, \quad \alpha\to 0, \quad \text{and} \quad   \frac{\alpha}{\beta} \to\text{constant} .
\end{equation}
Unless otherwise stated, the variable $\theta$ runs in the interval $-\pi < \theta \leq \pi$.
 
\noindent  We use the Euler-MacLaurin  summation formula: 
\begin{equation}
\label{Eq:sum_formula}
\sum_{n=n_{\text{min}}}^{\infty} f(x) = \int_{n_{\text{min}}}^{\infty}dx f(x) +\frac{1}{2} f(n_{\text{min}}) + \Delta.
\end{equation}
The rest term  $\Delta$ in the above formula is given by:
\begin{equation}
\label{Eq:sum_EM}
\Delta = -\sum_{l=1}^{\infty}  \frac{B_{2l}}{2l!} f^{(2l-1)}(n_{\text{min}}),
\end{equation}
where the $B_{2 l}$ are the Bernoulli numbers.
We will also use the Abel-Plana formulation  \cite{Dowling89sumformula} for the sum (\ref{Eq:sum_formula}) with $n_{min}=0$. In Abel-Plana formula the rest $\Delta$ is expressed as:
\begin{equation}
\label{Eq:sum_AP}
\Delta = i \int_{0}^{\infty}\frac{f(i y)-f(-iy)}{e^{2\pi y}-1}. 
\end{equation}

\subsection{Asymptotics of $\gamma_{1}(\theta)$} 
From the  Eq.~\eqref{Eq:sum_formula}  and  Eq.~\eqref{Eq:sum_EM}) one obtains:
\begin{equation}
\gamma_{1}(\theta) = \int_{1}^{\infty}dx f_1(x) +\frac{\sin (\theta)}{2 \sinh (\beta)} -\sum_{l=1}^{\infty}  \frac{B_{2l}}{2l!} f_{1}^{(2l-1)}(1).
\label{Eq:EMgamma1}
\end{equation}
The above expression can be rewritten in the form:
\begin{equation}
\gamma_{1}(\theta) =\frac{1}{\beta}\left(\int_{0}^{\infty} dx -\int_{0}^{\beta} dx\right)\frac{\sin (x \theta/\beta)}{\sinh (x)}  +\frac{\sin (\theta)}{2 \sinh (\beta)} -\sum_{l=1}^{\infty}  \frac{B_{2l}}{2l!} f_{1}^{(2l-1)}(1).
\end{equation}
In the limit $\beta \to 0$ we can use the expansion $\sinh{\beta n} = \beta n +O(\beta^2)$: the dominant contribution $\Delta^{0}_1$ in the expansion of the rest term:
\begin{equation}
\sum_{l=1}^{\infty}  \frac{B_{2l}}{2l!} f_{1}^{(2l-1)}(1)= -\beta^{-1}\Delta^{0}_1+ O(\beta^0),
 \end{equation} 
 is given by:
 \begin{equation}
 \label{Eq:rest_1}
 \Delta^{0}_1 = -\sum_{l=1}^{\infty}  \frac{B_{2l}}{2l!} \frac{d^{2l-1}}{dx^{2l-1}}\left[\frac{\sin{(x \theta)}}{x}\right](1)
 \end{equation}
The integrals:
 \begin{equation}
\int_{0}^{\infty} dx\; \frac{\sin (x y)}{\sinh (x)} =\frac{\pi}{2}\tanh{\left( \frac{\pi}{2}y\right)}, \quad \text{Si}(\theta)\equiv \int_{0}^{1} \;dx \frac{\sin{(\theta)x}}{x} , 
\end{equation} 
give, for the expansion (\ref{Eq:EMgamma1}), the following result:
\begin{equation}
\gamma_{1}(\theta)  = \frac{1}{\beta}\left[ \frac{\pi}{2}\tanh{\left(\frac{\pi \theta}{2 \beta}\right)} +\frac{\sin (\theta)}{2}- \text{Si}(\theta)+ \Delta_{1}^{0}\right] + O(\beta^0).
\end{equation}
The term $\Delta_{1}^{0}$ in (\ref{Eq:rest_1}) can also be computed by applying the Euler-Maclaurin formula to the sum:
\begin{equation}
\sum_{n=1}^{\infty} \frac{\sin{\left( n \theta \right)}}{n} = 
\begin{cases}
    \frac{\pi}{2}-\frac{\theta}{2},& \text{if }\; 0<\theta \leq \pi\\
   -\frac{\pi}{2}-\frac{\theta}{2},              & \text{if} \;\pi\leq \theta < 0\\
   0              & \text{if} \;\theta = 0,
\end{cases}.
\end{equation}
One obtains:
\begin{equation}
\Delta_{1}^{0} =  \frac{\theta}{2}-\frac{\sin (\theta)}{2}+ \text{Si}(\theta).
\end{equation}

\begin{figure}
\label{gamma1comp}
\begin{tikzpicture}
\draw (0,0) node[left]{\includegraphics[scale=0.25]{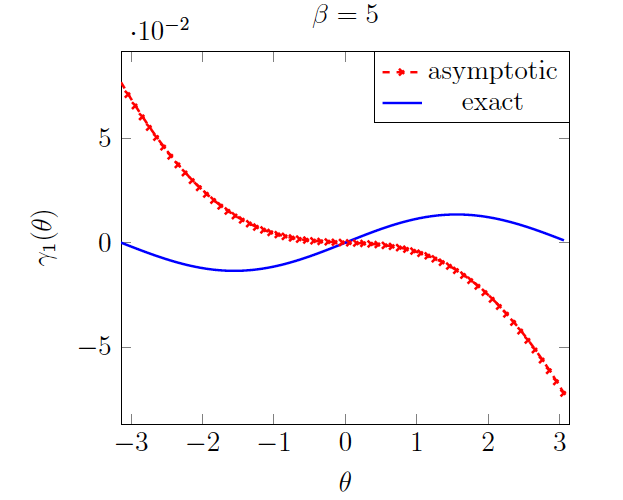}};
\draw (3,0) node{\includegraphics[scale=0.25]{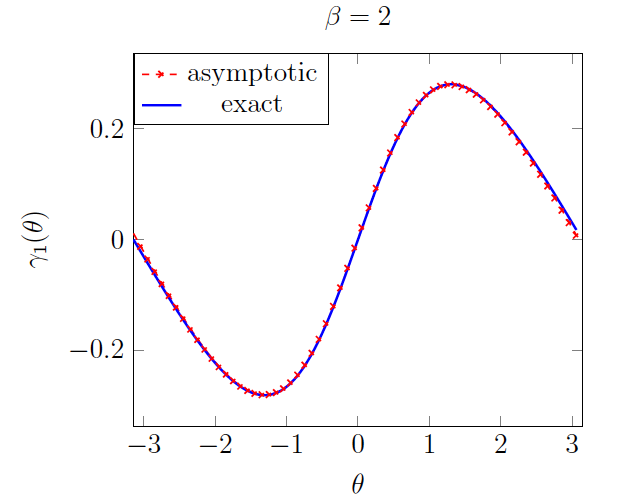}};
\draw (6,0) node[right]{\includegraphics[scale=0.25]{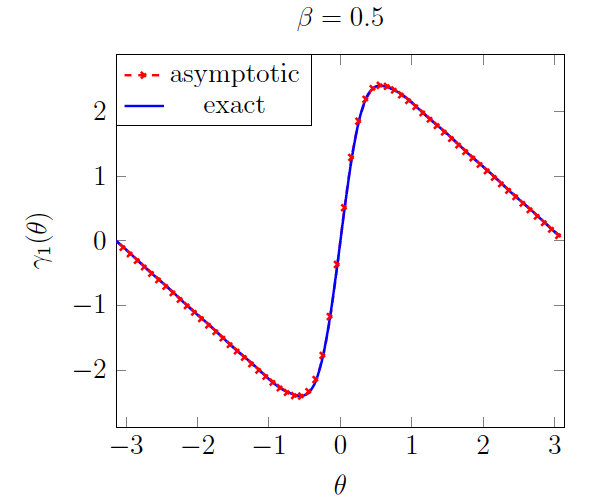}};
\end{tikzpicture}
\caption{The plots compare, for different values of $\beta= 5,2,0.5$, the function $\gamma_{1}(\theta)$ computed using the sum in (\ref{Eq:definitions}) and the asymptotic formula (\ref{Eq:gamma1_largeL})}
\end{figure}

\noindent  Collecting all the previous formula one finally gets:
\begin{equation}
\label{Eq:gamma1_largeL}
\gamma_{1}(\theta) = \frac{1}{\beta}\left(\frac{\pi}{2}\tanh{\left( \frac{\pi \theta}{2 \beta}\right)} -\frac{\theta}{2}\right)+O(\beta^0), \quad \text{for}\; -\pi < \theta \leq \pi
\end{equation}
The validity of (\ref{Eq:gamma1_largeL}) is verified in Fig (\ref{gamma1comp}).
In the interval $0\leq \theta <2\pi$, the result (\ref{Eq:gamma1_largeL}) reads:
\begin{equation}
\label{Eq:gamma1_largeL_02pi}
\gamma_{1}(\theta) = \begin{cases}
    \frac{1}{\beta}\left(\frac{\pi}{2}\tanh{\left( \frac{\pi \theta}{2 \beta}\right)} -\frac{\theta}{2}\right),& \text{if }\; 0\leq\theta \leq \pi\\
   \frac{1}{\beta}\left(\frac{\pi}{2}\tanh{\left( \frac{\pi (\theta-2\pi)}{2 \beta}\right)} -\frac{\theta}{2}+\pi\right).              & \text{if} \;\pi< \theta < 2\pi
\end{cases} +O(\beta^0).
\end{equation}
The above formuala can be approximated (with a difference $\sim e^{-1/\beta}$) by the  expression:
\begin{equation}
\label{Eq:gamma1_largeL_th}
\gamma_{1}(\theta) = \frac{1}{\beta}\left(\frac{\pi}{2}-\frac{\theta}{2}+\frac{\pi}{2}\tanh{\left( \frac{\pi \theta}{2 \beta}\right)} +\frac{\pi}{2}\tanh{\left( \frac{\pi (\theta-2\pi)}{2 \beta}\right)} \right)+O(\beta^0)
\end{equation}

\subsection{Asymptotics of $\gamma_{2}(\theta)$} 

Using the identity:
\begin{equation}
\label{Eq:new_sum}
\sum_{n\in \mathbb{Z}^{+}} \frac{e^{i n \theta}}{\sinh{(\beta n -\alpha)}} = e^{i \theta+ \alpha} \sum_{p=0}^{\infty} \frac{e^{2\alpha p}}{e^{\beta(2p+1)}-e^{i \theta}},
\end{equation}
the sum $\gamma_2(\theta)$ can be written in the form:
\begin{equation}
\gamma_2(\theta) = -\frac{1}{2 \sinh{(\alpha)}}+e^{i \theta+ \alpha} \sum_{p=0}^{\infty} \frac{e^{2\alpha p}}{e^{\beta(2p+1)}-e^{i \theta}}-e^{-i \theta- \alpha} \sum_{p=0}^{\infty} \frac{e^{-2\alpha p}}{e^{\beta(2p+1)-e^{-i \theta}}}
\end{equation}

\noindent Applying the summation formula (\ref{Eq:sum_formula}) to the (\ref{Eq:new_sum}) and using the integral representation of the hypergeometric functions $_2 F_1 (a,b;c;z)$, we obtain:
\begin{eqnarray} 
\gamma_{2}(\theta)&=&-\frac{1}{2 \sinh{(\alpha)}}+e^{\alpha+i \theta} \left[\frac{e^{-\beta}}{2 (\beta-\alpha)} \, _2F_1\left(1,1-\frac{\alpha}{\beta};2-\frac{\alpha}{\beta};e^{i \theta-\beta}\right)+\frac{1}{2 (e^{\beta}-e^{i \theta})}\right]-\nonumber \\
&& -e^{-\alpha-i \theta} \left[\frac{e^{-\beta}}{2 (\beta+\alpha)} \, _2F_1\left(1,\frac{\alpha}{\beta}+1;\frac{\alpha}{\beta}+2;e^{-\beta-i \theta}\right)+\frac{1}{2 (e^{\beta}-e^{-i \theta})}\right]+\Delta_{2},
\end{eqnarray}
where a convenient expression for the rest $\Delta_{2}$ is found by the formula (\ref{Eq:sum_AP}):
\begin{equation}
\Delta_{2} = i \int_{0}^{\infty} dy  \frac{h^{+}(i y) -h^{-}(i y)-h^{+}(-i y)+ h^{-}(-i y)}{e^{2\pi y}-1}, \quad h^{\pm}(x) = \frac{e^{\pm 2 \alpha x }}{e^{\beta(1+2x)-e^{\pm i \theta}}}.
\end{equation}

\subsubsection{$\beta, \alpha \to 0$ and $\theta>>\alpha,\beta$}
In the limit $\beta, \alpha \to 0$ and $\theta>>\alpha,\beta$, it is straightforward to obtain from the above expression that:
\begin{equation}
\gamma_{2}(\theta)=\beta^{-1}\hat{\gamma}_{2}(\theta)+O(\alpha^0, \beta^0),
\end{equation}
where
\begin{equation}
\label{Eq:hat_gamma} 
\hat{\gamma}_{2}(\theta)=-\frac{\beta}{2\alpha}+e^{i \theta} \left[\frac{\beta}{2 (\beta-\alpha)} \, _2F_1\left(1,1-\frac{\alpha}{\beta};2-\frac{\alpha}{\beta};e^{i \theta}\right)\right]-e^{-i \theta} \left[\frac{\beta}{2 (\beta+\alpha)} \, _2F_1\left(1,\frac{\alpha}{\beta}+1;\frac{\alpha}{\beta}+2;e^{-i \theta}\right)\right].
\end{equation}
The above expression expresses the leading behavior of $\gamma_{2}(\theta)$ for values of $\theta >>\alpha, \beta$. 
More subtle is to find the contribution to the leading term of $\gamma_{2}(\theta)$ coming from the region $\theta \leq \alpha,\beta$. In order to derive this contribution, we computed the small $\theta$ expansion of $\gamma_{2}(\theta)$.

\subsubsection{$\beta, \alpha \to 0$ and $\theta <<\beta, \alpha$}
Differently from what seen before, we use the summation formula (\ref{Eq:sum_formula}) and (\ref{Eq:sum_EM}) on the sum $\gamma_2(\theta)$ as expressed in (\ref{Eq:definitions}):
\begin{equation}
\label{Eq:gamma2_EM2}
\gamma_{2}(\theta)=\frac{1}{2}\left(\int_{2}^{\infty} dy \; f_{2}^{+}(x)-\int_{1}^{\infty} dy \;f_{2}^{-}(x)\right) +\frac{e^{ i \theta}}{2\sinh{(\beta-\alpha)}}-\frac{1}{2\sinh{\alpha}}+\Delta_2,
\end{equation}
where the rest $\Delta_{2}$ is expressed now as:
\begin{equation}
\Delta_{2} =\sum_{l=1}^{\infty} \frac{B_{2l}}{(2l)!} \left(\frac{d^{(2 l -1)}}{d x^{2l-1}}f^{-}_2(1)-\frac{d^{(2 l -1)}}{d x^{2l-1}}f^{+}_2(2)\right).
\end{equation}
It is convenient to express the functions under consideration in terms of  the parameters $(\tilde{\theta}, \beta, \tilde{\varepsilon})$ that are defined as: 
\begin{equation}
\tilde{\epsilon}= 2 \frac{\alpha}{\beta}-1,\quad \tilde{\theta} = \frac{\theta}{\beta}.
\end{equation}
The contribution coming from the integrals in (\ref{Eq:gamma2_EM2}) admits the following expansion in $\beta$ : 
\begin{eqnarray}
&&\frac{1}{2}\left(\int_{2}^{\infty} dy \; f_{2}^{+}(x)-\int_{1}^{\infty} dy \;f_{2}^{-}(x)\right) =\frac{e^{i \tilde{\theta} \beta(1-\tilde{\epsilon})/2}}{2 \beta}\left( \int_{\beta(3/2-\tilde{\epsilon}/2)}^{\infty}\; dx \;\frac{e^{i x \tilde{\theta} }}{\sinh{(x)}}-\int_{\beta(3/2+\tilde{\epsilon}/2)}^{\infty}\; d x\;\frac{e^{-i x \tilde{\theta} }}{\sinh{(x)}}\right)=\nonumber \\
&& = \beta^{-1}\left[- \text{coth}^{-1}\left(\frac{ \tilde{\epsilon}}{3}\right) +i\frac{\pi}{2} \tanh\left(\frac{\pi \tilde{\theta}}{2}\right)-i\frac{\pi}{2}\right]+ O(\beta^0).
\end{eqnarray}
The rest $\Delta_{2}$ behaves as:
\begin{equation}
\Delta_{2} = \beta^{-1} \Delta^{0}_{2}(\tilde{\epsilon})+ O(\beta^{0}, \tilde{\theta}),
\end{equation}
where $\Delta^{0}_{2}(\tilde{\epsilon})$ is some function of $\tilde{\epsilon}$. We obtain therefore the following result:
\begin{equation}
\label{Eq:gamma2_exp_beta}
\gamma_2 (\tilde{\theta}) = \beta^{-1}\left[-\text{coth}^{-1}\left(\frac{ \tilde{\epsilon}}{3}\right) +i\frac{\pi}{2}\tanh\left(\frac{\pi \tilde{\theta}}{2}\right)-i\frac{\pi}{2} +\frac{1}{-1 + \tilde{\epsilon}}+ \frac{1}{1 + \tilde{\epsilon}}+\Delta^{0}_{2}(\tilde{\epsilon})\right] +O(\beta^0)
\end{equation}
In the small $\tilde{\epsilon}$ expansion, the  (\ref{Eq:gamma2_exp_beta}) yields:
\begin{equation}
\gamma_2(0) = \beta^{-1}\left[d_0 + \tilde{\epsilon}\left(-\frac{7}{3}+d_1\right) + O(\tilde{\epsilon}^2)\right] +O(\beta^0),
\end{equation}
where $d_0$ and $d_1$  are the coefficients of the expansion of $\Delta^{0}_{2}(\tilde{\epsilon})=d_0+ d_1 \tilde{\epsilon}+O(\epsilon^2)$. The direct computation of $\gamma_2(0)$:
\begin{eqnarray}
\gamma_2(0) &=&\frac{1}{\beta} \sum_{n\geq 0}\left[\frac{1}{2n+(1+\tilde{\epsilon})}-\frac{1}{2n+(1-\tilde{\epsilon})}\right]+O(\beta^0)=\beta^{-1} \left[ -2 \tilde{\epsilon}\sum_{n\geq 0}\frac{1}{(2n+1)^2} +O(\tilde{\epsilon})\right]+O(\beta^0)=\nonumber \\ 
&=& \beta^{-1} \left[-\frac{\pi^2}{4}\tilde{\epsilon}+O(\tilde{\epsilon}^2)\right] +O(\beta^0),
\end{eqnarray}
fixes $d_0=0$ and $d_1=-\pi^2/4+7/3$.

\begin{figure}
\label{gamma2comp}
\begin{tikzpicture}
\draw (0,0) node[left]{\includegraphics[scale=0.25]{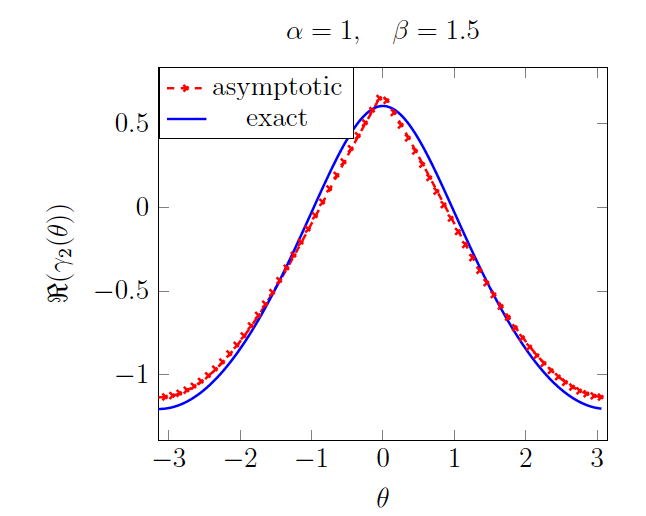}};
\draw (0,-5) node[left]{\includegraphics[scale=0.25]{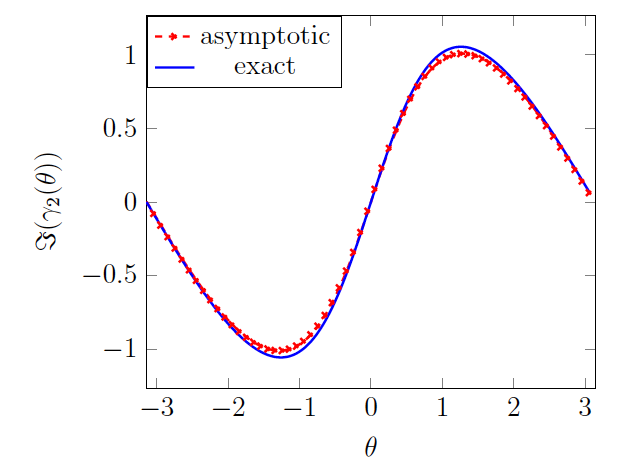}};
\draw (3,0) node{\includegraphics[scale=0.25]{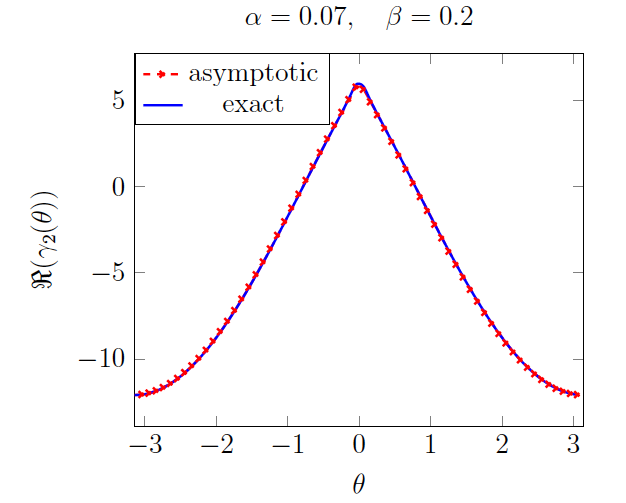}};
\draw (3,-5) node{\includegraphics[scale=0.25]{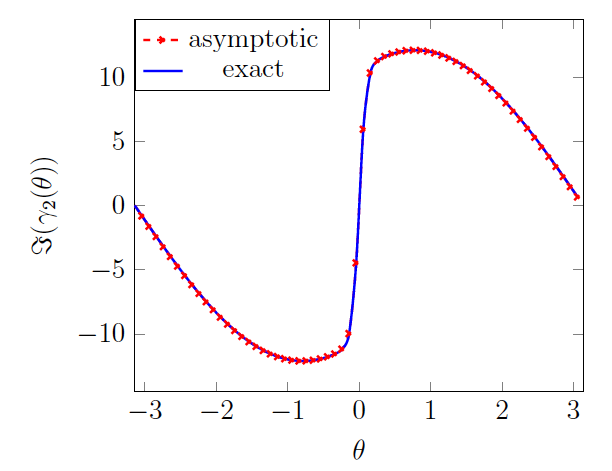}};
\end{tikzpicture}
\caption{The plots compare the real and imaginary part of the functions $\gamma_{2}(\theta)$ computed from  (\ref{Eq:definitions}) and its asymptotics (\ref{Eq:gamma2_largeL}) for values of $(\alpha,\beta)= (1,1.5), (0.1,0.15)$. }
\end{figure}

\noindent We have obtained therefore the following expansion of $\gamma_{2}(\theta)$ for small $\tilde{\epsilon}$:
\begin{equation}
\label{Eq: gamma2_smalltheta}
\gamma_2(\tilde{\theta})= \beta^{-1}\left[-\frac{\pi^2}{4} \tilde{\epsilon}+ i \frac{\pi}{2} \tanh\left(\frac{\pi\tilde{\theta}}{2}\right)+O(\tilde{\epsilon}^2)\right]+O(\beta^0).
\end{equation}
Comparing the above result with the small $\tilde{\epsilon}$ expansion of  $\hat{\gamma}_{2}(0)$ in (\ref{Eq:hat_gamma}):
\begin{equation}
 \hat{\gamma}_{2}(0) =\begin{cases}
    -i \frac{\pi}{2} - \frac{\pi^2}{4} \tilde{\epsilon}+O(\tilde{\epsilon}^2),& \text{if }\; 0\leq\theta \leq \pi\\
  i \frac{\pi}{2} - \frac{\pi^2}{4} \tilde{\epsilon}+O(\tilde{\epsilon}^2),              & \text{if} \;-\pi < \theta < 0
\end{cases},
\end{equation}
we find:
\begin{equation}
\label{Eq:gamma2_largeL}
\gamma_{2}(\theta) = \begin{cases}
    \beta^{-1}\left[\hat{\gamma}_{2}(\theta) + i \frac{\pi}{2}\left(\tanh\left(\frac{\pi \theta/\beta}{2}\right)-1\right)\right],& \text{if }\; 0\leq\theta \leq \pi\\
   \beta^{-1}\left[\hat{\gamma}_{2}(\theta) + i \frac{\pi}{2}\left(\tanh\left(\frac{\pi \theta/\beta}{2}\right)+1\right)\right].              & \text{if} \;-\pi< \theta < 0
\end{cases} +O(\beta^0). 
\end{equation}
The above asymptotics are checked  in Fig. (\ref{gamma2comp}).
A compact expression  in the interval $0\leq \theta<2\pi$  providing an approximation $\sim e^{-1/\beta}$ to the asymptotics (\ref{Eq:gamma2_largeL}) is:

\begin{equation}
\label{Eq:gamma2_largeL_th}
\gamma_{2}(\theta)=\beta^{-1}\left[\hat{\gamma}_{2}(\theta) +i\frac{\pi}{2}\left(\tanh{\left( \frac{\pi \theta}{2 \beta}\right)} +\tanh{\left( \frac{\pi (\theta-2\pi)}{2 \beta}\right)}\right)\right]+O(\beta^0)
\end{equation}

\subsection{Asymptotics of $\gamma_{3}(\theta)$} 

In the limit $\beta,\alpha\to 0$ and in the region  $\theta >> \beta,\alpha$, one can replaces $\cosh(\beta n -\alpha)$ by one, as  $\cosh(\beta n-\alpha)= 1+ O(\beta^{-2})$. 

In this case one obtains:
\begin{equation}
\gamma_3(\theta) = 0 + O(\beta^{-2})\quad \theta >> \beta
\end{equation}

\begin{figure}
\label{gamma3comp}
\begin{center}
\begin{tikzpicture}
\draw (0,0) node[left]{\includegraphics[scale=0.25]{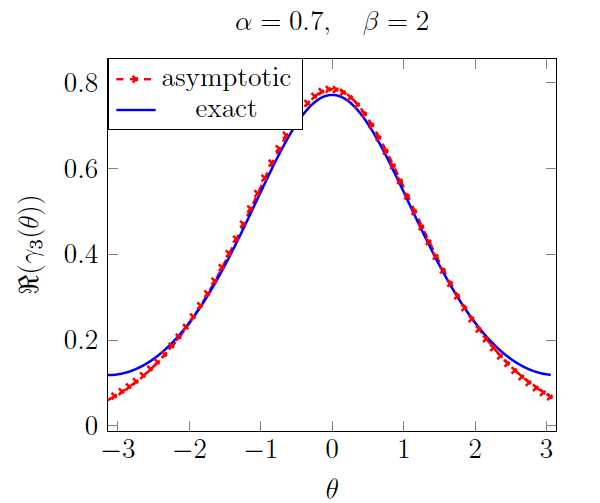}};
\draw (2.5,0) node[right]{\includegraphics[scale=0.25]{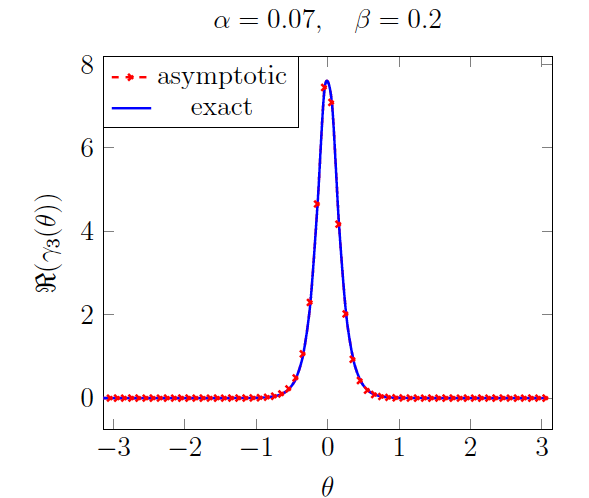}};
\draw (0,-5) node[left]{\includegraphics[scale=0.25]{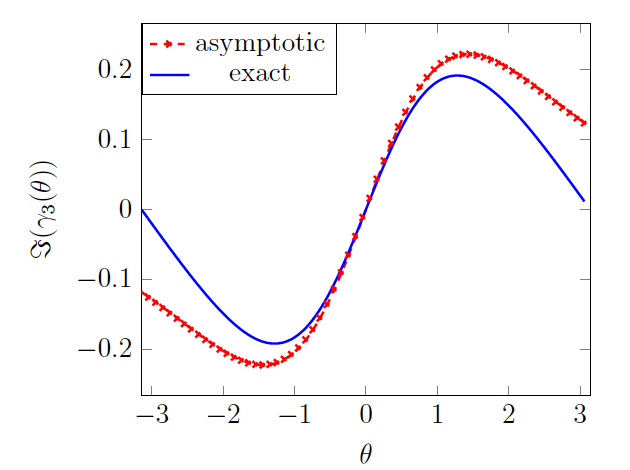}};
\draw (2.3,-5) node[right]{\includegraphics[scale=0.25]{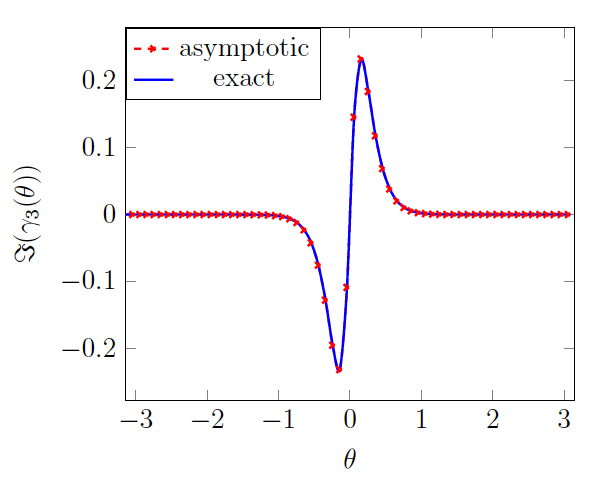}};
\end{tikzpicture}
\caption{The plots compares the real and imaginary  part of the function $\gamma_{3}(\theta)$ as in (\ref{Eq:definitions}) with its asymptotics (\ref{Eq:gamma3_largeL_th_1})}
\end{center}
\end{figure}

In the region $\theta \leq \beta$, the asymptotics of the function $\gamma_{3}(\theta)$  can be found as in the previous case by using the formula (\ref{Eq:sum_EM}):
\begin{equation}
\label{Eq:gamma3_EM2}
\gamma_{3}(\theta)=\frac{1}{2}\left(\int_{2}^{\infty} dy \; f_{3}^{+}(x)+\int_{1}^{\infty} dy \;f_{3}^{-}(x)\right) +\frac{e^{ i \theta}}{2\cosh{(\beta-\alpha)}}-\frac{1}{2\cosh{\alpha}}+\Delta_3
\end{equation}
where
\begin{equation}
\Delta_{3} =\sum_{l=1}^{\infty} \frac{B_{2l}}{(2l)!} \left(\frac{d^{(2 l -1)}}{d x^{2l-1}}f^{-}_3(1)-\frac{d^{(2 l -1)}}{d x^{2l-1}}f^{+}_3(2)\right).
\end{equation}
One can verify that the rest term $\Delta_3$ behaves as:
\begin{equation}
\Delta_3 = i\left( e^{i \theta}-e^{2 i \theta}\right) \frac{-2 + \theta \cot(\theta/2)}{2 x}+O(\beta^{2}),
\end{equation}
thus producing a term of order $\beta^0$, $\Delta_3 = O(1)$
The only terms that contribute with a $O(\beta^{-1})$ term are the  integrals in (\ref{Eq:gamma3_EM2}). Using the same manipulations as before, one obtains:
\begin{equation}
\label{Eq:gamma3_largeL_th_1}
\gamma_{3}(\theta)= \beta^{-1}\left[\frac{\pi}{2}e^{i \theta \frac{\alpha}{\beta}}\cosh\left(\frac{\pi \theta}{\beta 2}\right)^{-1}
\right] + O(\beta^0).
\end{equation}
The validity of the above asymptotic is tested  in Fig. (\ref{gamma3comp}).
In the interval $0\leq \theta <2\pi$ the (\ref{Eq:gamma3_largeL_th_1}) takes the form:
\begin{equation}
\label{Eq:gamma3_largeL_th}
\gamma_{3}(\theta) = \begin{cases}
    \beta^{-1}\left[\frac{\pi}{2}e^{i \theta \frac{\alpha}{\beta}}\cosh\left(\frac{\pi \theta}{\beta 2}\right)^{-1}
\right] ,& \text{if }\; 0\leq\theta \leq \pi\\
   \beta^{-1}\left[\frac{\pi}{2}e^{i (\theta-2\pi) \frac{\alpha}{\beta}}\cosh\left(\frac{\pi (\theta-2\pi)}{\beta 2}\right)^{-1}
\right] .              & \text{if} \;-\pi< \theta < 0
\end{cases} +O(\beta^0). 
\end{equation}

\bibliography{2D_bib}

\end{document}